\documentclass[longauth]{aa} 
\usepackage{graphicx}
\usepackage[varg]{txfonts}
\usepackage{longtable}
\usepackage{pdflscape}
\usepackage{placeins}
\usepackage{float}
\usepackage{multicol}
\usepackage{xcolor}
\usepackage{chemformula}
\usepackage{upgreek}
\usepackage{booktabs}

\usepackage[breaklinks=true]{hyperref} 
\usepackage{etoolbox}
\apptocmd{\thebibliography}{\setlength{\itemsep}{0pt}\setlength{\parskip}{0pt}}{}{}

\def\ms{\hbox{\,m\,s$^{-1}$}}         
\def\kms{\hbox{\,km\,s$^{-1}$}}       
\def\Msun{\hbox{$\mathrm{M}_{\odot}$}}           

\def\Mjup{\hbox{$\mathrm{M}_{\rm Jup}$}}        

\def\Me{\hbox{$\mathrm{M}_{\oplus}$}\,}             
\def\Re{\hbox{$\mathrm{R}_{\oplus}$}\,}
\def\mp{M_{\rm p}}
\def\rp{R_{\rm p}}

\newcommand{\lrhk}{\ensuremath{\log R'_{\rm HK}}}
\newcommand{\bis}{\ensuremath{{\rm BIS}}}

\definecolor{royalblue}{rgb}{0.255, 0.412, 0.882}

\newcommand{\micron}{$\upmu$m}

\begin{document}

\title{The GAPS programme at TNG\thanks{Based on: observations made with the Italian \textit{Telescopio Nazionale Galileo} (TNG), operated on the island of La Palma by the INAF - \textit{Fundaci\'on Galileo Galilei} at the \textit{Roque de Los Muchachos} Observatory of the \textit{Instituto de Astrof\'isica de Canarias} (IAC).}}
\subtitle{LXXI. A sub-Neptune suitable for atmospheric characterization in a multiplanet and mutually inclined system orbiting the bright K dwarf TOI-5789 (HIP\,99452)} 
\titlerunning{A transiting sub-Neptune in the multiple and mutually inclined system TOI-5789 (HIP\,99452)}
\authorrunning{Bonomo et al.}

\author{A.~S.~Bonomo\inst{\ref{oatorino}}
\and L.~Naponiello\inst{\ref{oatorino}}
\and A.~Sozzetti\inst{\ref{oatorino}}
\and S.~Benatti\inst{\ref{oapalermo}}
\and I.~Carleo\inst{\ref{oatorino}, \ref{iac}, \ref{ull}}
\and K.~Biazzo\inst{\ref{oaroma}}
\and P.~E.~Cubillos\inst{\ref{srigraz}}
\and M.~Damasso\inst{\ref{oatorino}}
\and C.~Di~Maio\inst{\ref{oapalermo}}
\and C.~Dorn\inst{\ref{ethzurich}}
\and N.~Hara\inst{\ref{lam}}
\and D.~Polychroni\inst{\ref{oatorino}}
\and M.-L.~Steinmeyer\inst{\ref{ethzurich}}
\and K.~A.~Collins\inst{\ref{cfa}}
\and S.~Desidera\inst{\ref{oapadova}}
\and X.~Dumusque\inst{\ref{unige}}
\and A.~F.~Lanza\inst{\ref{oacatania}}
\and B.~S.~Safonov\inst{\ref{sai}}
\and C.~Stockdale\inst{\ref{hazelobs}}
\and D.~Turrini\inst{\ref{oatorino}}
\and C.~Ziegler\inst{\ref{sassu}}
\and L.~Affer\inst{\ref{oapalermo}}
\and M.~D'Arpa\inst{\ref{oapalermo}}
\and V.~Fardella\inst{\ref{oapalermo}}
\and A.~Harutyunyan\inst{\ref{tng}}
\and V.~Lorenzi\inst{\ref{tng}}
\and L.~Malavolta\inst{\ref{oapadova}, \ref{unipadova}}
\and L.~Mancini\inst{\ref{unitor}, \ref{oatorino}}
\and G.~Mantovan\inst{\ref{unipadova}, \ref{oapadova}}
\and G.~Micela\inst{\ref{oapalermo}}
\and F.~Murgas\inst{\ref{iac}, \ref{ull}}
\and D.~Nardiello\inst{\ref{unipadova}, \ref{oapadova}}
\and I.~Pagano\inst{\ref{oacatania}}
\and E.~Pall\'e\inst{\ref{iac}, \ref{ull}}
\and M.~Pedani\inst{\ref{tng}}
\and M.~Pinamonti\inst{\ref{oatorino}}
\and M.~Rainer\inst{\ref{oabrera}}
\and G.~Scandariato\inst{\ref{oacatania}}
\and R.~Spinelli\inst{\ref{oapalermo}}
\and T.~Zingales\inst{\ref{unipadova}, \ref{oapadova}}
}

\institute{
INAF - Osservatorio Astrofisico di Torino, via Osservatorio 20, 10025 Pino Torinese, Italy \label{oatorino}
\and INAF - Osservatorio Astronomico di Palermo, Piazza del Parlamento 1, I-90134 Palermo, Italy \label{oapalermo}
\and Instituto de Astrof\'{i}sica de Canarias (IAC), 38205 La Laguna, Tenerife, Spain \label{iac}
\and Departamento de Astrof\'isica, Universidad de La Laguna (ULL), E-38206 La Laguna, Tenerife, Spain \label{ull}
\and INAF-Osservatorio Astronomico di Roma, Via Frascati 33, I-00040 Monte Porzio Catone (RM), Italy \label{oaroma}
\and Space Research Institute, Austrian Academy of Sciences, Schmiedlstrasse 6, A-8042, Graz, Austria \label{srigraz}
\and Institute for Particle Physics and Astrophysics, ETH Zürich, Otto-Stern-Weg 5, 8093 Zürich, Switzerland \label{ethzurich}
\and Universit\'e Aix Marseille, CNRS, CNES, LAM, Marseille, France \label{lam}
\and Center for Astrophysics \textbar \ Harvard \& Smithsonian, 60 Garden Street, Cambridge, MA 02138, USA \label{cfa}
\and INAF - Osservatorio Astronomico di Padova, Vicolo dell'Osservatorio 5, I-35122 Padova, Italy \label{oapadova}
\and D\'epartement d'astronomie de l'Universit\'e de Gen\`eve, Chemin Pegasi 51, 1290 Versoix, Switzerland \label{unige}
\and INAF - Osservatorio Astrofisico di Catania, Via S. Sofia 78, I-95123 Catania, Italy \label{oacatania}
\and Sternberg Astronomical Institute Lomonosov Moscow State University, Universitetskii prospekt, 13, Moscow, Russia \label{sai}
\and Hazelwood Observatory, Churchill, Victoria 3840, Australia 
\label{hazelobs}
\and Dept. of Physics, Engineering and Astronomy, Stephen F. Austin State Univ., 1936 North St, Nacogdoches, TX 75962, USA \label{sassu}
\and Fundaci{\'o}n Galileo Galilei - INAF, Rambla Jos{\'e} Ana Fernandez P{\'e}rez 7, E-38712 Bre$\tilde{\rm n}$a Baja (TF), Spain \label{tng}
\and Dipartimento di Fisica e Astronomia, Università degli Studi di Padova, Vicolo dell’Osservatorio 3, I-35122 Padova, Italy \label{unipadova}
\and Dipartimento di Fisica, Universit\`a di Roma ``Tor Vergata'', Via della Ricerca Scientifica 1, I-00133 Roma, Italy \label{unitor}
\and INAF – Osservatorio Astronomico di Brera, Via E. Bianchi 46, 23807 Merate (LC), Italy \label{oabrera}
}

\date{Received 12 October 2025 / Accepted 13 January 2026}

\offprints{\email{aldo.bonomo@inaf.it}}

\abstract{Sub-Neptunes with planetary radii of $\rp \simeq 2-4\,R_{\oplus}$ are the most common planets around solar-type stars in short-period ($P<100$~d) orbits. It is still unclear, however, what their most likely composition is, that is whether they are predominantly gas dwarfs or water worlds. The sub-Neptunes orbiting bright host stars are very valuable because they are suitable for atmospheric characterization, which can break the well-known degeneracy in planet composition from the planet bulk density, when combined with a precise and accurate mass measurement. 
Here we report on the characterization of the sub-Neptune TOI-5789\,c, which transits in front of the bright ($V=7.3$~mag and $K_{\rm s}=5.35$~mag) and magnetically inactive K1\,V dwarf HIP\,99452 every 12.93\,days, thanks to TESS photometry and 141 high-precision radial velocities obtained with the HARPS-N spectrograph. We find that its radius, mass, and bulk density are $R_{\rm c}=2.86^{+0.18}_{-0.15}~\rm R_\oplus$, $M_{\rm c}=5.00 \pm 0.50~\rm M_\oplus$, and  $\rho_{\rm c}=1.16 \pm 0.23$~g\,cm$^{-3}$, respectively, and we show that TOI-5789\,c is a  promising target for atmospheric characterization with both JWST and, in the future, Ariel. 
By analyzing the HARPS-N radial velocities with different tools, we also detected three additional non-transiting planets, namely TOI-5789\,b, d, and e, with orbital periods and minimum masses of 
$P_{\rm b}=2.76$\,d, $M_{\rm b}\sin{i}=2.12 \pm 0.28~\rm M_\oplus$, $P_{\rm d}=29.6$\,d, $M_{\rm d}\sin{i}=4.29 \pm 0.68~\rm M_\oplus$, and $P_{\rm e}=63.0$\,d, $M_{\rm e}\sin{i}=11.61 \pm 0.97~\rm M_\oplus$.
TOI-5789 is a mutually inclined system as the difference between the orbital inclinations of planets b and c must be higher than $\sim4$\,deg. Nevertheless, from sensitivity studies based on both the HARPS-N and archival HIRES radial-velocity measurements, we can exclude the possibility that these relatively high mutual inclinations are due to the perturbation by an outer gaseous giant planet.
}


 
\maketitle
\nolinenumbers

%

 \section{Introduction}
Sub-Neptunes are the most common planets around solar-type stars in short-period ($P<100$\,d) orbits, with an occurrence rate of $~\sim 40\%$ \citep{Fulton2017}. 
These planets have typical masses of $\mp \lesssim 20$\,\Me, and radii of $\rp \simeq 2-4$\,\Re above the well-known radius valley at $R_{\rm trans}\sim1.7-2.0$\,\Re. This valley is thought to separate prevalently rocky planets
($R_{\rm p} \lesssim R_{\rm trans}$) from non-rocky ones ($R_{\rm p} \gtrsim R_{\rm trans}$; \citealt{Fulton2017, VanEylen2018}). 

Despite their abundance, the most likely composition and nature of sub-Neptunes are not well known. They may be either gas dwarfs with a rocky core and a H$_2$-dominated atmosphere (e.g., \citealt{Bodenheimer2014}), or  water worlds with a water-rich (icy) interior surrounded by a steam atmosphere (e.g., \citealt{Zeng2019}). 
Another possible composition consists of a miscible envelope, where $\rm H_2/He$ are mixed with water vapor and other volatile species produced by chemical reactions between a magma-ocean and a hydrogen-rich envelope, on top of a rocky core  \citep{Benneke2024}. 

The gas dwarf (and miscible envelope) or water-world compositions would  originate from two very different formation  scenarios. Gas dwarfs are indeed expected to have formed within the water iceline (i.e., the water condensation front at $\sim1-3$\,au for solar-type stars) and they have possibly undergone a modest migration by accreting a primordial atmosphere with approximately the same metallicity as their host stars (e.g., \citealt{LeeChiang2016}). In contrast, water worlds are expected to have formed beyond the water iceline and undergone a substantial migration toward their star (e.g., \citealt{Izidoro2021}).

Several inferences on the typical sub-Neptune composition were made by attempting to reproduce the radius valley. While some studies seemed to strongly favor the gas dwarf composition coupled with atmospheric photoevaporation \citep{OwenWu2017, Modirrousta-Galian2020}, more recent works consider a mixed population of gas dwarfs and water-rich sub-Neptunes \citep{Burn2024, ChakrabartyMulders2024}. After all, the existence of water worlds is predicted by models of planet formation and migration (e.g., \citealt{Venturini2024}).

Some inferences on the composition of sub-Neptunes can also be made from their bulk densities, which are generally determined through space-based transit photometry and high-precision radial-velocity (RV) follow-up. Several sub-Neptunes around M dwarfs \citep{{LuquePalle2022}} and FGK dwarfs \citep{Bonomo2025a}
seem to have the icy compositions predicted for water worlds. However, alternative compositions with rocky interiors and (very) thin $\rm H_2/He$ envelopes cannot be excluded, given the well-known degeneracy between rocky, icy, and gaseous mass fractions that can match the same bulk density (e.g., \citealt{Rogers_2010, Spiegel2014, Neil2022}).

The characterization of the atmospheres of sub-Neptunes can  break the abovementioned composition degeneracy in the absence of thick clouds, provided that the planet mass can be determined to a certain precision and accuracy (e.g., \citealt{Batalha2019, DiMaio2023}). Indeed, water worlds should have metal-rich atmospheres, in particular for the expected higher abundance of water vapor, while gas dwarfs should possess solar-metallicity ones (e.g., \citealt{Miller-Ricci2009, Kempton2023}). 

In the present study, we report on the planetary nature confirmation and characterization of the  sub-Neptune transiting candidate TOI-5789.01, which was discovered  by the Transiting Exoplanet Survey Satellite (TESS) space mission \citep{Ricker2015, Sullivan2015} 
around the bright K dwarf HIP\,99452. 
The host star has a  stellar companion with a mass of about $0.24\,\Msun$ (spectral type M5\,V;  \citealt{Newton2014}) at $\sim103.9\arcsec$, which corresponds to a large projected separation of $\sim2120$\,au \citep{Sapozhnikov2020}. 
This makes TOI-5789 an S-type planetary system, even though the stellar companion may not have affected it, given the large distance.
Thanks to high-precision RV follow-up, we derived the planet mass and bulk density, and detected three low-mass and mutually inclined siblings. We finally show that the transiting planet is suitable for atmospheric characterization with transmission spectroscopy.

\section{Data}

\subsection{Photometry}

\subsubsection{TESS space-based photometry} 
\label{tess_photometry}
HIP\,99452 was observed by TESS in Sectors 54 (from July 9 to August 5, 2022) and 81 (from July 15 to August 10, 2024). The target was named TOI-5789 (TESS Object of Interest; \citealt{Guerrero2021}) on September 22, 2022, after two transit-like events were identified in Sector 54 Full-Frame Images (FFIs; \citealt{Ricker2015}) by the MIT Quick-Look Pipeline (QLP; \citealt{Huang2020}), which performs aperture photometry on FFIs and searches for periodic transits. In particular, TOI-5789.01 was flagged as a candidate planet with $P=12.9256\,\pm\,0.0023$\,days, a transit depth of $\sim 790$\,ppm (parts per million), and a corresponding radius of $2.54\,\pm\,0.15\,R_{\oplus}$. 

We inspected both the Presearch Data Conditioning Simple Aperture Photometry (PDC-SAP; \citealt{Stumpe2012,Stumpe2014}, \citealt{Smith2012}) and the SAP \citep{Twicken2010,Morris2020} light curves produced by the TESS Science Processing Operations Center (SPOC; \citealt{Jenkins2016}) pipeline at the NASA Ames Research Center, without finding any modulation related to stellar magnetic activity. In total, TESS recorded four transits of TOI-5789.01, at 10-minute cadence in the first sector (Fig.~\ref{fig:sector_54}) and both 2-minute and 20-second cadences in the second one (Fig.~\ref{fig:sector_81}), with the fastest cadence used for our analysis.

\subsubsection{Ground-based photometry} 
\label{ground_photometry}
We observed a full transit window of TOI-5789.01 on UTC 2024 October 07 in Pan-STARRS $z_s$ band from the Las Cumbres Observatory Global Telescope (LCOGT) \citep{Brown:2013} 0.35\,m network node at McDonald Observatory near Fort Davis, Texas, United States (McD). The 0.35\,m Planewave Delta Rho 350 telescopes are equipped with a $9576\times6388$ pixels QHY600 CMOS camera having an image scale of $0.73\arcsec$ per pixel, resulting in a $114\arcmin\times72\arcmin$ full field of view. The images were calibrated by the standard LCOGT {\tt BANZAI} pipeline \citep{McCully:2018}, and differential photometric data were extracted using {\tt AstroImageJ} \citep{Collins:2017}. We used circular $11\arcsec$ photometric apertures, which excludes any flux from all known Gaia DR3 catalog neighbors that are bright enough to possibly be the source of the TESS transit detection. 
We observed no significant flux decreases in any of the close stars at the predicted transit time.
We also checked the available ASAS-SN photometry \citep{Shappee2014, Kochanek2017} for any signals related to the stellar magnetic activity and, in particular, the stellar rotation period, but found none that were significant.

\subsection{High-angular resolution imaging}
High-angular resolution imaging is needed to search for nearby sources that can contaminate the TESS photometry, resulting in underestimated planetary radii, or be the source of astrophysical false positives, such as background eclipsing binaries. We searched for stellar companions to TOI-5789 with speckle imaging on the 4.1-m Southern Astrophysical Research (SOAR) telescope \citep{Tokovinin2018} on 4 November 2022 UT, observing in Cousins $I$-band, a similar visible bandpass as TESS. This observation was sensitive with $5\sigma$ detection to a 5.6-magnitude fainter star at an angular distance of $1\arcsec$ from the target. More details of the observations within the SOAR TESS survey are available in \citet{Ziegler2020}. The $5\sigma$ detection sensitivity and speckle auto-correlation functions from the observations are shown in Figure \ref{fig:det_sensitivity_imaging} (left panel). No nearby stars were detected within $3\arcsec$ of TOI-5789 in the SOAR observations.

TOI-5789 was also observed on August 28, 2023, with the speckle polarimeter on the 2.5-m telescope at the Caucasian Observatory of Sternberg Astronomical Institute (SAI) of Lomonosov Moscow State University. A low--noise CMOS detector Hamamatsu ORCA--quest \citep{Strakhov2023} was used. The atmospheric dispersion compensator was active. A medium band filter centered on 625~nm with 50~nm FWHM provided the angular resolution of $0.083\arcsec$. Long exposure atmospheric seeing was very good at the moment of observation: $0.63^{\prime\prime}$. No companion was detected. The detection limits at distances $0.25$ and $1.0^{\prime\prime}$ from the star are $\Delta=5.5^m$ and $8.4^m$ (see Fig.~\ref{fig:det_sensitivity_imaging}, right panel).

\subsection{Radial velocities}
\label{radial_velocities}
\subsubsection{HIRES archival radial velocities}
We downloaded 79 archival RVs  gathered with the HIRES (HIgh Resolution Echelle Spectrometer) spectrograph at the Keck telescope \citep{Howard2010} from \citet{Rosenthal2021}. We then removed two outliers identified with Chauvenet's criterion (e.g., \citealt{Bonomo2023}) at 2450276.99192 and 2452804.99617 $\rm BJD_{UTC}$. The HIRES RVs have a scatter of 3.57\,\ms, which is approximately 3 times higher than the median of the formal uncertainties, that is 1.26\,\ms. They are displayed in Fig.~\ref{fig:HARPSN_HIRES_RVs}.

\subsubsection{HARPS-N radial velocities}
We observed TOI-5789 with the HARPS-N (High Accuracy Radial velocity Planet Searcher in North hemisphere) spectrograph at the Telescopio Nazionale Galileo \citep{Cosentino2012} as part of different observing programs. In the framework of the GAPS2 project to characterize Neptune-size planets (PI: G. Micela; see, e.g., \citealt{Naponiello2022}), we collected 41 spectra between June and September 2023 with typical exposure times of 900 s and a median signal-to-noise ratio (S/N) of 207 at 5500~\AA. 
The observations continued from October 2023 to November 2024 within the ``Ariel Masses Survey (ArMS)'' large program\footnote{The five-year ArMS observing program is focused on the mass determination of small transiting planets with radii spanning the radius valley, which are suitable for atmospheric characterization with JWST \citep{Gardner2006} and, in the future, Ariel \citep{Tinetti2018}.} (ProgID: AOT48TAC\_48; PI: S. Benatti), 
which is also part of the GAPS collaboration. Specifically, we collected 79 additional HARPS-N spectra  with typical exposure times of 900 s and median S/N of 204. Finally, we also obtained 22 spectra, with typical exposure time of 300~s and median S/N of 124, through three observing programs on the TNG Spanish time between May 2023 and August 2024 (ProgIDs: CAT23A\_52, CAT23B\_74, CAT24B\_20; PI: I. Carleo).

We uniformly extracted the RVs from all these spectra using the updated ESPRESSO Data Reduction Software (DRS, v3.0.1), adapted for HARPS-N spectra \citep{Dumusque2021}. We used a K0V stellar template, containing 3842 lines weighted according to their relative depth, to compute the Cross-Correlation Function (CCF; see, e.g.,  \citealt{2002A&A...388..632P} and references therein), and provided a systemic RV of $-49.31$\kms as input to the pipeline. 

\begin{figure}[t!]
\hspace{-0.7 cm}
\includegraphics[width=8.0cm, angle=90]{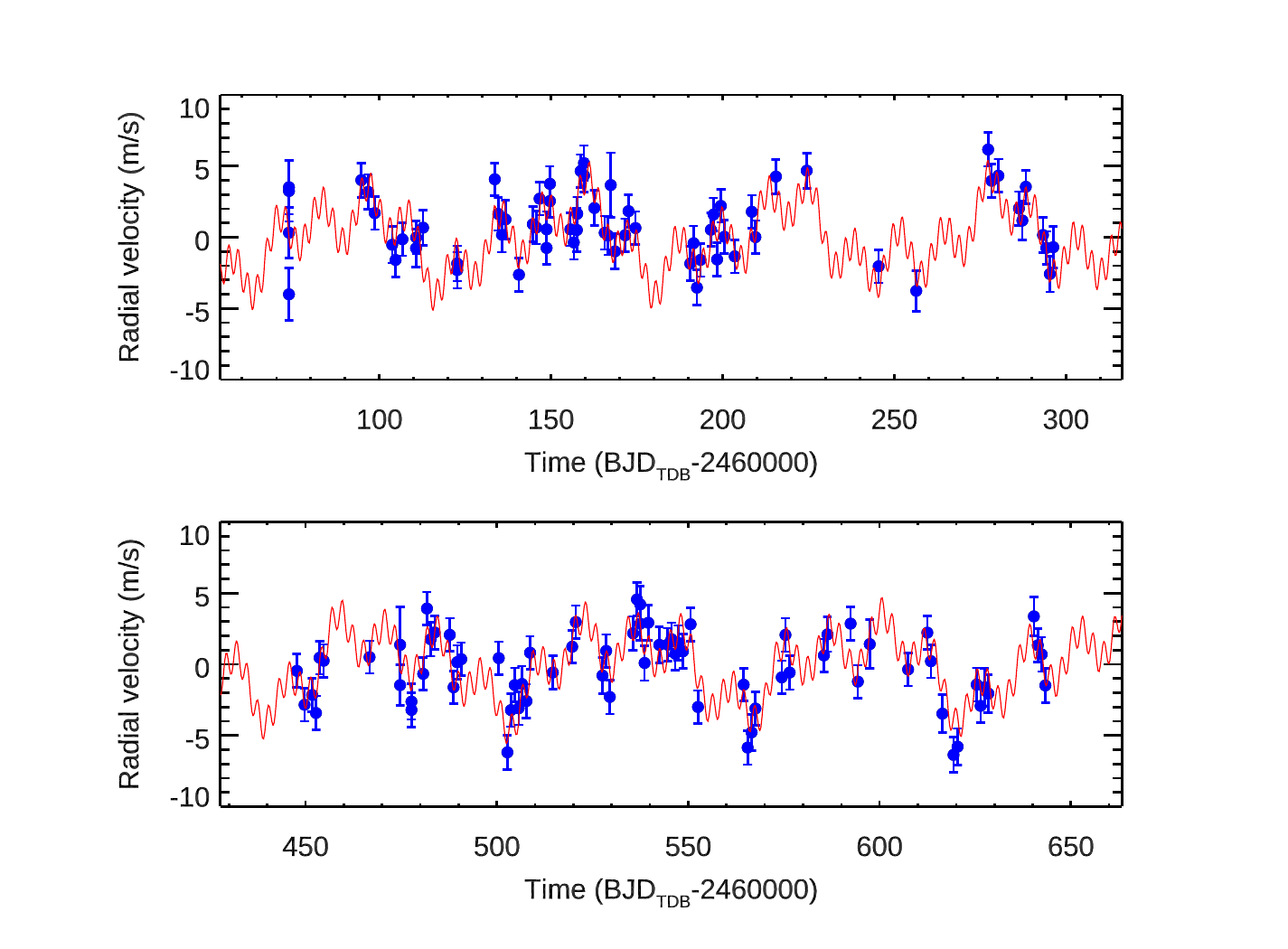}
\caption{HARPS-N radial velocities of TOI-5789 (blue circles) along with the four-planet Keplerian model (solid red line; see Sect.~\ref{RV_modeling}). The error bars take the RV jitter into account.}
\label{fig:HARPSN_RVs}
\end{figure}

We checked for possible outliers in the HARPS-N RVs by using Chauvenet's criterion, and found one at 2460213.4266 $\rm BJD_{TDB}$, which was then excluded from the analyses. The HARPS-N RVs have a scatter of 2.58\,\ms, which is almost seven times higher than the median of the formal uncertainties, namely 0.38\,\ms. They are shown in Fig.~\ref{fig:HARPSN_RVs}.

The HARPS-N pipeline also provides activity indicators that can be useful to check for possible stellar activity signals in our RV time series, such as the CCF bisector span (\bis), the CCF full width at half maximum (FWHM) and contrast, from which it is possible to compute the CCF equivalent width \citep{Klein2024},
and the  chromospheric CaII H\&K $S$-index and \lrhk  (see Fig.~\ref{fig:RVs_ActInd}). The latter has an average value of $-5.153 \pm 0.003$, indicating that the host star is magnetically inactive, and shows a low-amplitude long-term trend of unclear origin. There are no significant correlations between RVs and activity indices, being the Pearson coefficient $\left|r\right| < 0.20$.
The RVs and activity indicators are given in Table~\ref{tab:data}.

\section{Stellar parameters and elemental abundances}
\label{star_param}
We derived the stellar atmospheric parameters (effective temperature $T_{\rm eff}$, surface gravity $\log g$, microturbulence velocity $\xi$, and iron abundance [Fe/H]) by applying a method based on equivalent widths (EW) of Fe lines. We considered the same line list as in  \citet{Biazzoetal2022}, and measured the EWs using the ARES code (v2; \citealt{Sousaetal2015}). We then adopted the 1D LTE Kurucz model atmospheres linearly interpolated from the \cite{CastelliKurucz2003} grid with solar-scaled chemical composition and new opacities (ODFNEW). To derive the final parameters, we used the pyMOOGi code by \cite{Adamowetal2017}, which is a Python wrapper of the {\sc MOOG} code (\citealt{sneden1973}, version 2019). In particular, following \cite{Biazzoetal2022}, $T_{\rm eff}$ was derived by imposing the excitation equilibrium of Fe lines, $\log g$ through the ionization equilibrium of \ion{Fe}{i} and \ion{Fe}{ii} lines, and $\xi$ by removing the trend between the \ion{Fe}{i} abundances and the reduced equivalent width ($EW/\lambda$). Once the stellar atmospheric parameters and iron abundance were computed, we also measured the abundances of magnesium ([Mg/H]) and silicon ([Si/H]), considering the same line list as above. The derived stellar properties are reported in Table\,\ref{tab:star}.

\begin{figure}
\centering
\includegraphics[width=1\linewidth]{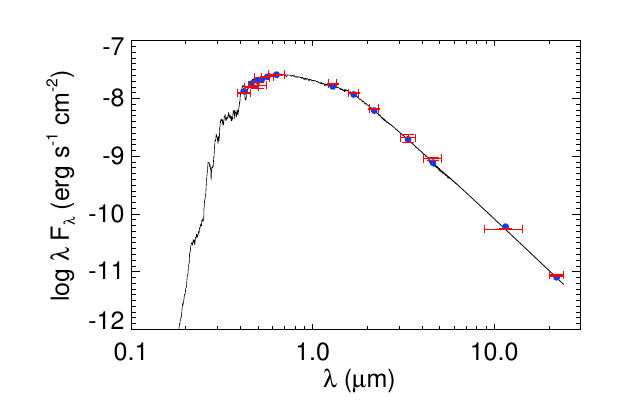}
\caption{Stellar spectral energy distribution. The broadband measurements from the Tycho, APASS Johnson and Sloan, 2MASS and WISE magnitudes are shown in red, and the corresponding theoretical values with blue circles. The solid black line displays the non-averaged best-fit model.} 
\label{fig:stellarSED}
\end{figure}

To determine the stellar mass, radius, and age, we simultaneously modeled the stellar Spectral Energy Distribution (SED) and the MIST evolutionary tracks (e.g., \citealt{Paxton2015}) in a differential evolution Markov chain Monte Carlo (DE-MCMC) Bayesian framework through the {\tt EXOFASTv2} tool (\citealt{2017ascl.soft10003E, Eastman2019}; see also \citealt{Naponiello2025} for more details). To this end, we imposed Gaussian priors on the Gaia DR3 parallax as well as on the $T_{\rm eff}$ and [Fe/H] derived from our previous analysis of HARPS-N spectra. To fit the SED, we used the Tycho-2 $B_{\rm T}$ and $V_{\rm T}$, APASS Johnson $B$ and $V$, Sloan $g'$ and $r'$ magnitudes, the 2MASS near-infrared $J$, $H$, and $K_{\rm s}$ magnitudes, and the WISE $W1$, $W2$, $W3$, and $W4$ infrared magnitudes (Table~\ref{tab:star}). The stellar SED and its best fit are shown in Fig.~\ref{fig:stellarSED}. From the medians and 15.86\%-84.14\% percentiles of the posteriors of the stellar mass, radius, and age, we derive $M_\star=0.821^{+0.032}_{-0.027}~\rm M_\odot$, $R_\star=0.833 \pm 0.023~\rm R_\odot$, and age $t=9.4^{+2.0}_{-3.5}$~Gyr. The old age is fully consistent with the low magnetic activity level of the host star.

The chromospheric activity index $\lrhk$ is close to the basal level, indicating a very old star with an age greater than $9$~Gyr according to both \cite{MamajekHillenbrand08} and \cite{Carvalho-Silvaetal25}. By extrapolating the empirical relation between \lrhk and stellar rotation periods ($P_{\rm rot}$) \citep{Noyes1984, MamajekHillenbrand08} to the inactive regime at $\lrhk < -5.0$, we find $P_{\rm rot} = 52 \pm 2$~d for the stellar $B-V=0.856$.

We applied the method of \cite{JohnsonSoderblom87} to compute the Galactic spatial velocity components $U, V, W$, using the stellar coordinates, parallax, and proper motion components in Table~\ref{tab:star}, and the systemic velocity $\gamma$ in Table~\ref{tab:planet_parameters}. From the derived Galactic velocity components (Table~\ref{tab:data}), a kinematical age of $12.7^{+1}_{-3}$~Gyr can be estimated according to \cite{Almeida-FernandesRocha-Pinto18}, although kinematic age estimates may be inaccurate for individual stars.   

Applying the prescriptions by \citet{Bensby2014}, we obtain for TOI-5789 a thick-to-thin disk probability ratio of TD/D$\sim$20.6. Moreover, the target shows
a maximum height above the Galactic plane of 1.5 kpc \citep{Mackereth2018} and a [Mg/Fe] abundance ratio  
of 0.17$\pm$0.10 (Table~\ref{tab:star}). These values, together with the relatively low iron abundance, are compatible with a thick disk candidate, which in turn is consistent with the old stellar age.

\begin{table}[ht!]
\centering\renewcommand{\arraystretch}{1.3} %
\caption{Stellar parameters.} %
\label{tab:star} %
\resizebox{\hsize}{!}{
\begin{tabular}{l c  c}
\hline %
\hline  
\emph{\textbf{Parameter}} & Value & Source \\
\hline\hline  
\multicolumn{1}{l}{\large{Identifiers}} \\ 
\hline
TOI    & 5789 & TOI catalog \\
TIC    &  87216634 & TIC \\
HIP    &  99452 & HIP \\
Tycho-2     &  1618-00827-1 & Tycho-2 \\
2MASS    & J20110609+1611162 & 2MASS \\
{\it Gaia}    & 1809360187275432832 & {\it Gaia}~DR3 \\ 
\hline
\multicolumn{1}{l}{\large{Coordinates, motion, and distance}} \\ 
\hline
Right ascension $\alpha$\,(J2000.0) [h] & 20:11:06.07  & {\it Gaia}~DR3 \\
Declination $\delta$\,(J2000.0)  [deg] & +16:11:16.79  & {\it Gaia}~DR3 \\
Parallax $\pi$   [mas] & $48.9262 \pm 0.0230$ & {\it Gaia}~DR3 \\
Proper motion $\mu_\alpha \cos{\delta}$   [mas\,yr$^{-1}$]  & $-415.062\pm0.020$ & {\it Gaia}~DR3 \\
Proper motion $\mu_\delta$   [mas\,yr$^{-1}$]  & $398.487\pm0.015$ & {\it Gaia}~DR3 \\
Galactic velocity $U^{(1)}$    [km\,s$^{-1}$] & $31.376 \pm 0.0026$ & This work \\ 
Galactic velocity $V^{(1)}$    [km\,s$^{-1}$] & $-26.883 \pm 0.0065$ & This work \\ 
Galactic velocity $W^{(1)}$   [km\,s$^{-1}$] & $61.910 \pm 0.0254$ & This work \\ 
Distance $d$   [pc]  & $20.4392\pm0.0095$ & This work \\ 
\hline
\multicolumn{1}{l}{\large{Magnitudes}} \\ [2pt] %
\hline
$B_{\rm T}$     & $8.373\pm0.017$ & Tycho-2 \\ 
$V_{\rm T}$     & $7.422\pm 0.011$ & Tycho-2 \\ 
$B$     & $8.155\pm0.017$ & APASS Johnson \\ 
$V$     & $7.299\pm 0.006$ & APASS Johnson \\ 
$g'$     & $7.825\pm0.118$ & APASS Sloan \\
$r'$     & $7.064\pm0.023$ & APASS Sloan \\
$G$     & $7.1130\pm0.0028$ & {\it Gaia}~DR3 \\
$J$     & $5.826\pm0.030$ & 2MASS \\
$H$     & $5.421\pm0.021$ & 2MASS \\
$K_{\rm S}$      & $5.350\pm0.024$ & 2MASS \\ 
$W1$     & $5.296\pm0.166$ & AllWISE \\
$W2$     & $5.218\pm0.068$ & AllWISE \\     
$W3$     & $5.341\pm0.015$ & AllWISE \\ 
$W4$     & $5.285\pm0.032$ & AllWISE \\ 
$A_{\rm V}$     & $<0.057$ & This work$^{(2)}$ \\ 
\hline
\multicolumn{1}{l}{\large{Stellar parameters}} \\ 
\hline
Spectral type& K1\,V &  \\
Luminosity $L_{\star}$  [$L_{\sun}$] & $0.469\pm0.017$ & This work$^{(2)}$ \\
Mass $M_{\star}$   [$M_{\sun}$] & $0.821^{+0.032}_{-0.027}$ & This work$^{(2)}$ \\
Radius $R_{\star}$    [$R_{\sun}$] & $0.833\pm0.023$ & This work$^{(2)}$ \\
Effective temperature $T_{\rm eff}$ [K] & $5185\pm80$ & This work$^{(3)}$\\ 
Metallicity $\rm{[Fe/H]}$  [dex] & $-0.12\pm0.08$ & This work$^{(3)}$ \\
Magnesium abundance $\rm{[Mg/H]}$  [dex] & $0.05\pm0.07$ & This work$^{(3)}$ \\
Silicon abundance $\rm{[Si/H]}$  [dex] & $-0.03\pm0.07$ & This work$^{(3)}$ \\
Micro-turbulent velocity $V_{\rm mic}$   [km\,s$^{-1}$] & $0.97\pm0.04$ & This work$^{(3)}$ \\
Surface gravity $\log g_{\star}$   [cgs] & $4.51\pm0.15$ & This work$^{(3)}$ \\
Surface gravity $\log g_{\star}$   [cgs] & $4.511^{+0.026}_{-0.024}$ & This work$^{(2)}$ \\
Density $\rho_{\star}$  [g\,cm$^{-3}$] & $2.00^{+0.17}_{-0.15}$ & This work$^{(2)}$ \\
Activity index $\lrhk$  [dex] &  $-5.153\pm0.003$ & This work$^{(3)}$ \\
Age $t$   [Gyr] & $9.4^{+2.0}_{-3.5}$ & This work$^{(2)}$ \\
\hline %
\end{tabular} 
}
\tiny
\tablebib{\small TESS Primary Mission TOI catalog \citep{Guerrero2021}; TIC \citep{Stassun2018,Stassun2019}; HIP \citep{Perryman1997}; Tycho-2 \citep{hog}; APASS \citep{Henden2016}; {\it Gaia} DR3 \citep{Gaia2023}; 2MASS \citep{Cutri2003}; AllWISE \citep{Cutri2013}.}
\begin{flushleft}
\footnotemark[1]{\small Galactic velocity components, where $U$ is positive toward the Galactic anticenter, $V$ in the direction of the Galactic rotation, and $W$  toward the Galactic north pole}. \\
\footnotemark[2]{\small From the {\tt EXOFASTv2} modeling.} \\
\footnotemark[3]{\small From the HARPS-N spectral analysis.} \\
\end{flushleft}
\end{table}

\normalsize

\section{Data analysis and results} 
\label{data_analysis} 

\subsection{Analysis of radial velocities}

\subsubsection{Generalized Lomb Scargle periodograms}
\label{RV_periodograms}
We first searched for periodic signals in both the HARPS-N and HIRES RV time series with the Generalized Lomb-Scargle (GLS) periodograms \citep{Zechmeister2009} by using a quite stringent threshold of $10^{-5}$ for the theoretical False Alarm Probability (FAP). We find four highly significant signals ($\rm FAP < 10^{-8}$) in the HARPS-N data and their residuals, with periods of $63.4\pm0.5$, $12.95\pm0.02$, $2.763\pm0.001$, and $29.8\pm0.2$\,days, in descending order of significance (see Fig.~\ref{fig:GLS_periodograms}). No significant peaks at the abovementioned periods, and in general at periods shorter than $\sim 150$\,d, are found in the GLS periodograms of the activity indicators, as expected from the very low level of the host star's magnetic activity. 
This supports the hypothesis of the planetary origin of the signals seen in the RVs.

The 30\,d signal cannot be related to the Moon either, for several reasons. Firstly, the host star systemic velocity of $\gamma \simeq -49.3\,\kms$ is well outside the [-30, 30]\,\kms, where the CCF of the Moon might affect the stellar CCF by overlapping with it (e.g., \citealt{Bonomo2010, Malavolta2017}). Secondly, the contrast of the Moon CCF is expected to be negligible compared to that of the stellar CCF, given the brightness of the host star. Thirdly, there are no peaks in the spectral window at the frequency corresponding to the Moon synodic orbital period, which might cause aliases in the RV time series (see Fig.~\ref{fig:GLS_periodograms}, bottom panel).

Consequently, we attribute the detected signals with periods of $2.76$, $12.9$, $29.8$, and 63\,d to the planets TOI-5789\,b, c, d, and e. The second periodicity agrees within $1\sigma$ with the transiting period of the candidate TOI-5789.01, and can thus naturally be ascribed to the stellar wobble caused by the transiting planet TOI-5789\,c.

We detect no clear periodicity in the HIRES data likely due to a combination of a smaller number of RVs, lower cadence, and higher scatter (see Fig.~\ref{fig:HARPSN_HIRES_RVs}). This is reminiscent of the case of Kepler-10 \citep{Bonomo2025a}.

\subsubsection{Stacked Bayesian Generalized Lomb Scargle periodograms}
Since the period of TOI-5789\,d might be close to the rotation period of an old star such as TOI-5789, we assessed its significance as a function of time, as tracked by the increasing number of measurements, with a stacked Bayesian Generalized Lomb-Scargle periodogram\footnote{\url{https://anneliesmortier.wordpress.com/sbgls/}} (BGLS, \citealt{bgls2015A&A...573A.101M,sbgls2017A&A...601A.110M}).
Specifically, we ran it on the HARPS-N RV residuals, after removing the signals with periods of $2.76$, $12.9$, and 63\,d with a GLS pre-whitening.
As shown in Fig.~\ref{fig:sbgls}, the signal appears stable in frequency after $\sim$90 measurements  and statistically significant. At the end of the time series, its power  noticeably increases over that of its 1-year alias, which corresponds to $P\sim27.6$\,d.

\subsubsection{Apodized sine periodograms}
To further test whether the four RV signals are of planetary origin, we employed the method of~\citet{gregory2016} and \citet{hara2022b}, which consists in modeling the data with periodic Keplerian orbits, multiplied by the so-called apodization factor acting as a temporal window. We used a Gaussian temporal profile with a free mean and free variance, representing the time at which the window is centered, and the timescale for which the signal is non zero. If, for a periodic signal at a given frequency, the preferred timescale of the apodization is greater than the timespan of the observations, it is indicative of a purely periodic signal, which supports the planetary origin of the signal. In our analysis presented in Appendix~\ref{app:asp}, we find that the stability of the 2.76, 12.9, 29.8, and 63\,d signals is strongly supported by the data.

\subsection{Bayesian modeling of radial velocities}
\label{RV_modeling}
\subsubsection{DE-MCMC modeling}
\label{DE-MCMC_analysis}
For the modeling of RVs, we only considered the HARPS-N data, because the addition of the HIRES ones does not yield any improvement in the orbital solution. This is actually expected from i) the higher HIRES scatter, and ii) the non-detection of significant signals in the HIRES RVs (Sect.~\ref{RV_periodograms}).

We modeled the HARPS-N RVs with non-interacting Keplerians in the same DE-MCMC framework as \citet{Bonomo2023, Bonomo2025a}, by varying the number of planets from two to four, with the simplest two-planet model containing only TOI-5789\,c and e. For each planet, we fit five free parameters, that is the orbital period $P$; 
the inferior conjunction time $T_{\rm c}$, which is equivalent to the mid-transit time for  planet c; 
$\sqrt{e}\cos(\omega)$ and $\sqrt{e}\sin(\omega)$, where $e$ and $\omega$ are the orbital eccentricity and the argument of periastron;
and the RV semiamplitude $K$. We also included in our model the HARPS-N systemic velocity (or zero point) $\gamma$ and the jitter term $\sigma_{\rm RV,jit}$; the latter allows us to account for additional uncorrelated noise of unknown origin (instrumental, stellar, undetected planets, etc.). 

We used the priors as given in Table~\ref{tab:priors_parameters}, that is uniform priors for most of the free parameters, with the exception of $T_{\rm c,c}$ and $P_{\rm c}$, for which we adopted Gaussian priors from the transit fitting, and planet eccentricities. Specifically, we fixed the eccentricity of TOI-5789\,b to zero because its expected orbit circularization time\footnote{This was estimated from Eq.~6 of \citet{Matsumura2008} by assuming the planet radius to be equal to $0.9~\rm R_\oplus$, and the planet and stellar modified tidal quality factors to be $Q^{'}_{\rm p}=10^3$ and $Q^{'}_{\rm s}=10^6$, respectively.} of $\sim 400$~Myr is much lower than the stellar age ($\sim9.4$\,Gyr). We instead adopted half-Gaussian priors centered on zero with $\sigma_{\rm e}=0.098$ on the eccentricities of TOI-5789\,c, d, and e from \citet{VanEylen2019}, to avoid spurious high eccentricities, and hence unphysical dynamical instabilities due to orbit crossings (e.g., \citealt{Zakamska2011, Hara2019}). The solution from the RV fit is practically identical to that derived from the combined analysis of TESS transits and HARPS-N RVs (see Table~\ref{tab:planet_parameters} and Sect.~\ref{photRV_modeling}).

The Bayesian Information Criterion (BIC; see, e.g., \citealt{Burnham_Anderson_2004, Liddle2007}) computed for the two-planet, three-planet, and four-planet models clearly disfavors the former one, and favors the four-planet model over the three-planet one, being $BIC_{\rm 2pl}-BIC_{\rm 3pl}=22.8$ and 
$BIC_{\rm 3pl}-BIC_{\rm 4pl}=12.2$. The latter is higher than the threshold of $\Delta BIC=10$ usually considered for very strong evidence for a more complex model \citep{Kass_Raftery_1995}, that is the four-planet model in our case.

\begin{figure}[t!]
\centering
\includegraphics[width=9.0cm]{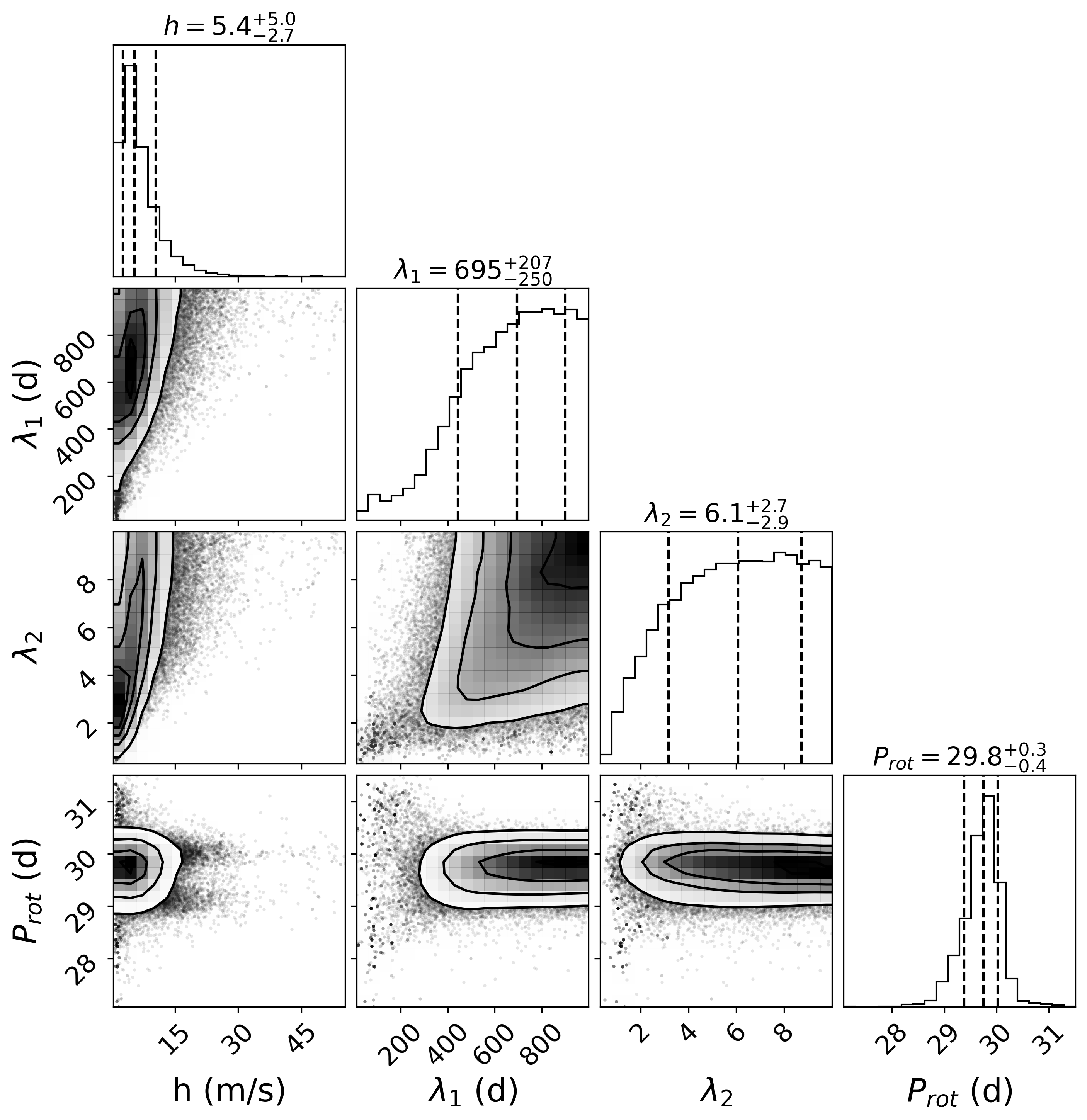}
\caption{Corner plot of the posterior distributions of the Gaussian process hyper-parameters. The posteriors are obtained with a three-planet model and Gaussian process regression to account for the 29.8\,d period. The posteriors of the parameters of the three planets are not shown for simplicity, but are fully consistent with those given in Table~\ref{tab:planet_parameters}.}
\label{fig:corner_plot}
\end{figure}

As mentioned in Sect.~\ref{RV_periodograms},  
no significant periodicity at $\sim 30$\,d is seen in any of the activity indicators, which would support the planetary origin of the 30\,d RV signal. Nonetheless, since this signal is close to a possible rotation period of magnetically quiet stars, we also modeled it with Gaussian process (GP) regression with a quasi-periodic kernel, following \citet{Bonomo2023} (see their Eq.~1). This GP modeling yielded RV semiamplitudes of planets b, c, and e fully consistent with those obtained with the four-planet model. In particular, for the transiting planet TOI-5789\,c, we get $K_{\rm c}=1.62\pm0.16$\,\ms and $1.56 \pm 0.15$\,\ms with the GP and non-GP modeling, respectively. Even more interesting, the GP exponential decay timescale ($\lambda_1$) and inverse harmonic complexity ($\lambda_2$) hyper-parameters both converge to the upper bounds of the wide intervals of their uniform priors (see Fig.~\ref{fig:corner_plot} and Table~\ref{tab:priors_parameters}). This indicates that this signal is both stable over the long term (long $\lambda_1$) and (quasi-)sinusoidal (large $\lambda_2$), as expected for a planetary signal. 
On the contrary, activity signals are typically characterized by $\lambda_1 \lesssim 150$\,d (e.g., \citealt{Giles2017, Bonomo2023}) and $\lambda_2 \lesssim 1.0$ (e.g., \citealt{Rajpaul2015, Bonomo2023}).

\subsubsection{Assessing the number of planets with \texttt{kima}}
We used \texttt{kima} \citep{faria2018} to make an independent blind inference on the number $N_{\rm p}$ of Keplerian signals detected in our RV dataset. \texttt{Kima} treats $N_{\rm p}$ as a free parameter and estimates its posterior distribution by determining the most probable number of Keplerians present in the data through a Bayesian model comparison.
To keep the test unbiased, we searched for up to ten Keplerian signals with the built-in function \texttt{RVmodel}, using 50000 steps, uncorrelated jitter, and the default priors for the orbital parameters of all the signals, with the exception of the eccentricity, for which we used the same truncated normal prior as in Sect.~\ref{DE-MCMC_analysis}. We did not use informative priors for the transiting planet c. 

\begin{figure}
    \centering
\includegraphics[width=0.5\textwidth]{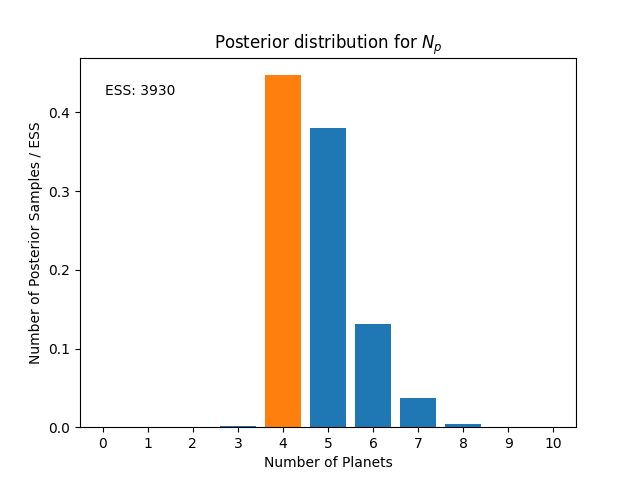}
    \caption{Posterior distribution of the number of planets $N_{\rm p}$ detected with \texttt{kima} in the HARPS-N radial velocities. The distributions are shown as the number of posterior samples for each value of $N_{\rm p}$ normalized to the effective sample size (ESS). The orange bar highlights the highest value  for $N_{\rm p}$=4 corresponding to the most likely 4-planet model.}
    \label{fig:kima}
\end{figure}

We find that the most probable number of signals in the data is $N_{\rm p}=4$ (Fig. \ref{fig:kima}), with orbital periods corresponding to those of the four signals discussed above. The probability ratio between the model with $N_{\rm p}=4$ and $N_{\rm p}=3$ is 352, which is larger than the threshold of 150 commonly adopted for the inclusion of an additional planet \citep[e.g.][]{faria2016A&A...588A..31F}, and is of the same order of the probability ratio from the $\Delta BIC$ (Sect.~\ref{DE-MCMC_analysis}), namely $\exp{(-0.5 \cdot \Delta BIC)}=446$ \citep{Burnham_Anderson_2004}. In contrast, all the models with $N_{\rm p}>4$ have probability ratios less than 1 compared to $N_{\rm p}=4$. The four Keplerian signals are recovered with  semiamplitudes in very good agreement with those found with the DE-MCMC analysis. 
Interestingly, \texttt{kima} blindly recovers the signal due to the transiting planet very well.

\subsection{Search for additional planets in TESS photometry}
\label{transit_search}
We analyzed both TESS sectors for additional transits using the Cambridge Exoplanet Transit Recovery Algorithm (CETRA; \citealt{Smith2025}), a fast and sensitive GPU-optimized transit detection tool. After masking the known TOI-5789\,c transits, we conducted a targeted search for signals with a minimum depth of 50 parts per million (ppm). This search yielded no reliable result, even at the shortest period of TOI-5789\,b, where a signal should have been readily detected. In particular, the transits of TOI-5789\,d should be visible in both sectors, and hence their non-detection confirms that the planet is not transiting. TOI-5789\,e is predicted to transit near the end of Sector 54, unless its transit occurs more than $1.6\sigma$ later than expected, in which case it would fall outside the TESS coverage. Therefore, a degree of ambiguity on the transiting nature of planet e still remains, which could be cleared out with additional photometry, for instance with CHEOPS (Characterising Exoplanet
Satellite; \citealt{Benz2021}).

In order to quantify the detectability of planet b, we performed a transit injection and recovery experiment using the \texttt{MATRIX} tool \citep{Pajares2023}. In particular, we masked the transits of TOI-5789\,c and detrended the light curve using the bi-weight approach and a window size of 0.5 days. We then injected a series of synthetic transits into the PDC-SAP light curve around the planet's orbital period, covering a range of planetary radii from $R_{\rm p}=0.5\,R_{\oplus}$ to $2.5\,R_{\oplus}$ with steps of $0.1\,R_{\oplus}$. Recovery was defined by successful retrieval of transit signals with duration within 1 hour of the set value, period within 5\% of the injected period, and a signal detection efficiency above 5.

The results indicate that only transits corresponding to planetary radii exceeding $R_{\rm p}\gtrsim0.9\,R_{\oplus}$ produce reliably recoverable signals. Planets smaller than this threshold fall below the detection limit of our pipeline, implying that a non-detection is consistent with $R_{\rm p}<0.9\,R_{\oplus}$. We conclude that either the orbit of TOI-5789\,b is  inclined with respect to TOI-5789\,c, or TOI-5789\,b is sufficiently small to escape photometric detection. However, the former hypothesis is much more likely than the latter, because a radius $R_{\rm p}<0.9\,R_{\oplus}$ for a mass of $M_{\rm p}\geq 2.12\,M_{\oplus}$ (as derived from the RVs; see Table~\ref{tab:planet_parameters}) would imply an unrealistic bulk density three times higher than that of the Earth. 

Finally, we searched for transit-time variations (TTVs) in the four available transits of TOI-5789\,c. Given the small number of events, we do not detect any significant departure from the predicted ephemeris within the current uncertainties.

\subsection{Combined modeling of TESS photometry and HARPS-N radial velocities}
\label{photRV_modeling}
We performed a combined modeling of the four TOI-5789\,c (PDC-SAP) TESS transits and HARPS-N RVs with two frameworks: the same DE-MCMC employed for the RV modeling (Sect.~\ref{RV_modeling})  \citep{Bonomo2015, Bonomo2017a}, and the Dynamic Nested Sampling through the use of \texttt{dynesty} \citep{Speagle2020}, as implemented in \texttt{juliet}\footnote{\texttt{\url{https://juliet.readthedocs.io/en/latest/}}} \citep{Espinoza2019}. 
In addition to the RV parameters  (Sect.~\ref{RV_modeling}), in the former framework we fit for the transit duration from first to fourth contact $T_{\rm 14}$;
the ratio of the planetary-to-stellar radii $R_{\rm p}/R_{*}$;
the inclination $i$ between the orbital plane and the plane of the sky;
the two limb-darkening coefficients $q_{1}=(u_{a}+u_{b})^2$ and $q_{2}=0.5 u_{a} / (u_{a}+u_{b})$ \citep{Kipping2013}, 
where $u_{\rm a}$ and $u_{\rm b}$ 
are the coefficients of the limb-darkening quadratic law in the TESS bandpass.
In the second framework, we instead fit $R_{\rm p}/R_{*}$ and the impact parameter $b$ through the $(r_1,r_2)$ parametrization described in \citet{Espinoza2018}. 
We oversampled the transit model by a factor of 30 in the first TESS sector to match the same fast cadence (20~s) as the second one.
We imposed uniform priors on all the transit parameters (Table~\ref{tab:priors_parameters}), except for a Gaussian prior on the stellar density from our determination in Sect.~\ref{star_param}, which affects all the transit parameters except $T_{\rm c}$ (cf. \citealt{Singh2022}).

\begin{figure*}[t!]
\centering
\includegraphics[width=8.5 cm, angle=180]{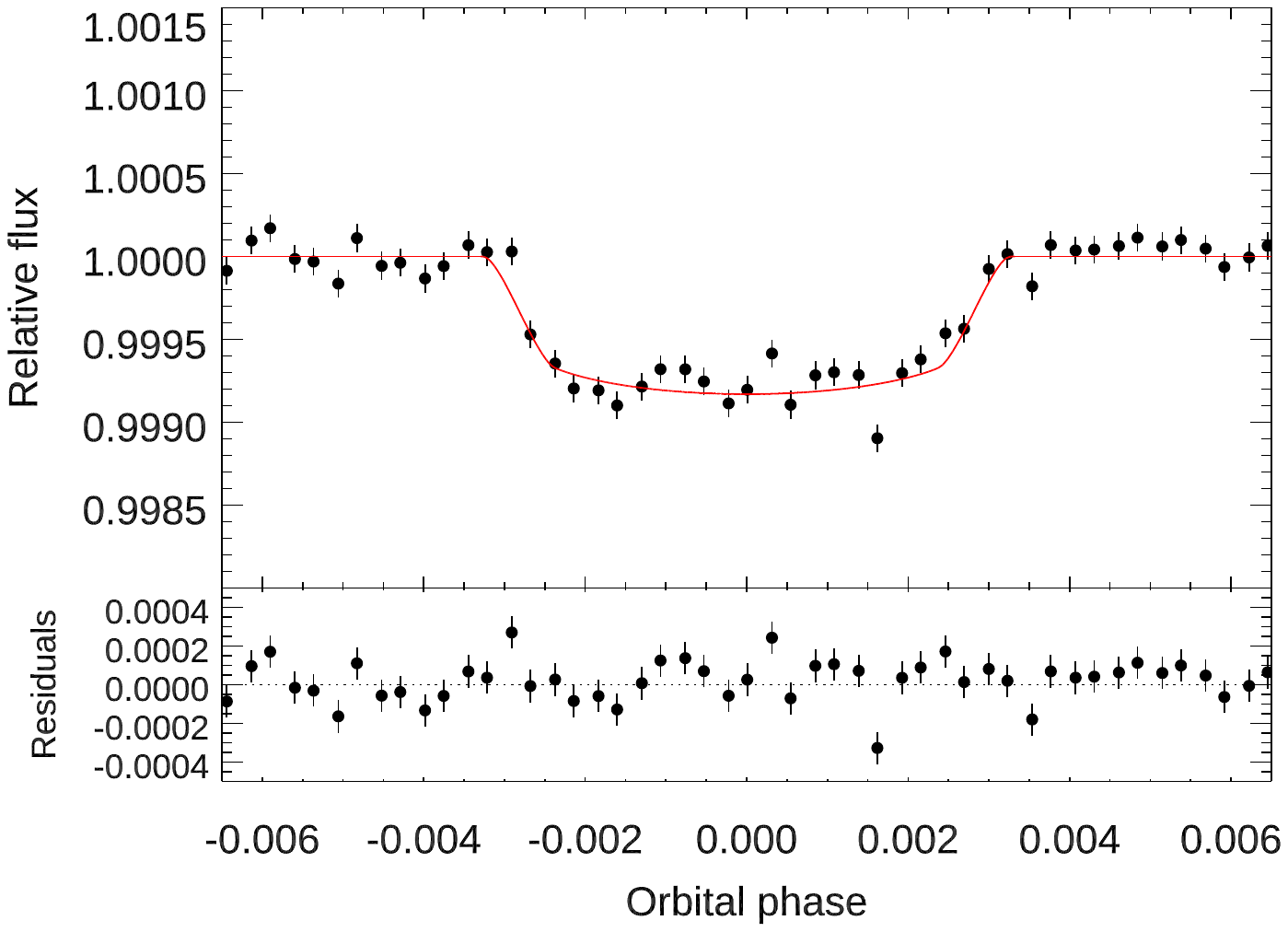}
\includegraphics[width=8.5 cm, angle=180]{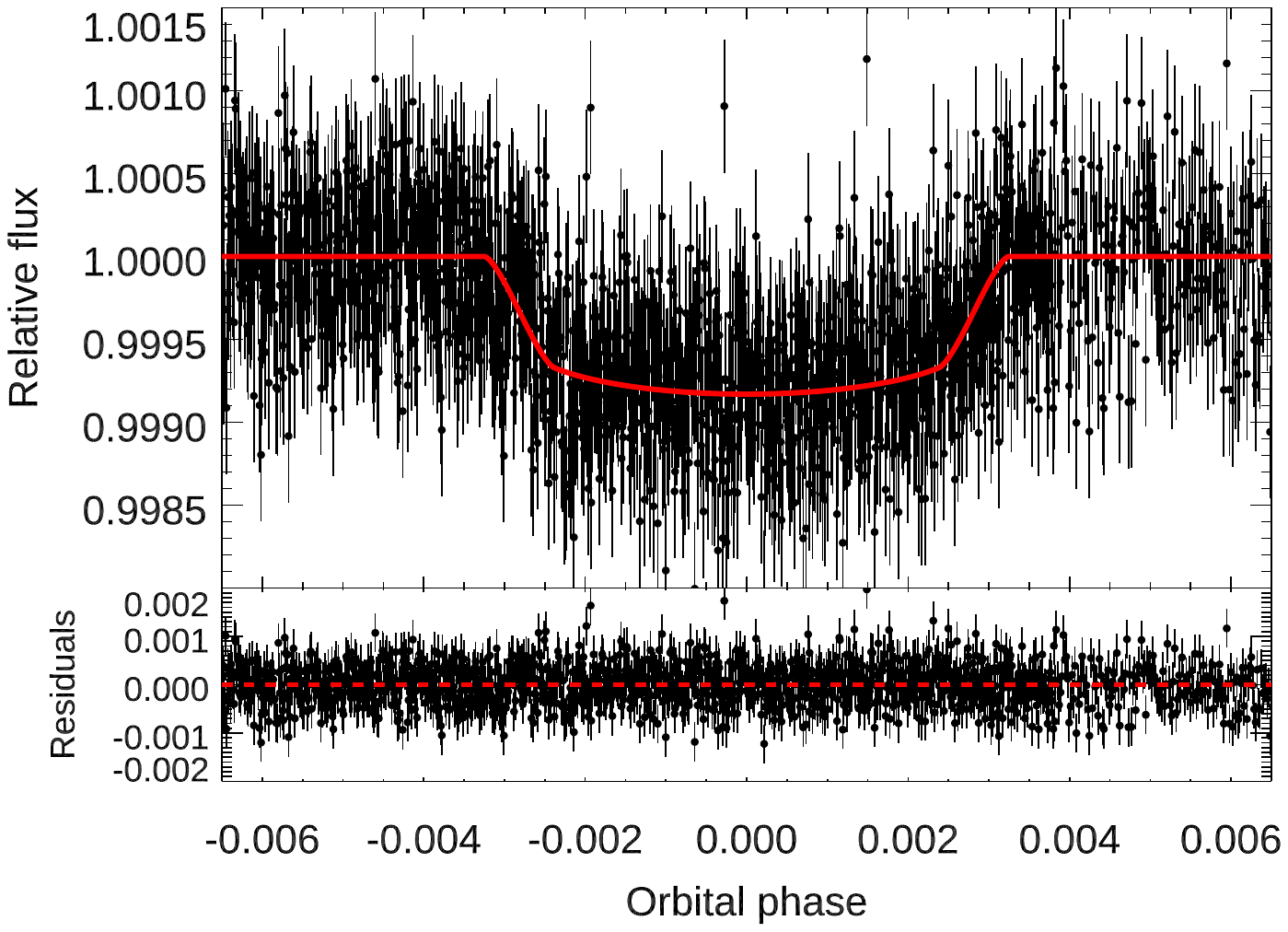} \\
\caption{Phase-folded transits of TOI-5789\,c with TESS long-cadence (left panel) and fast-cadence (right panel) data.}
\label{fig:transits_phase}
\end{figure*}

\begin{figure*}[h!]
\centering
\includegraphics[width=5.5 cm, angle=90]{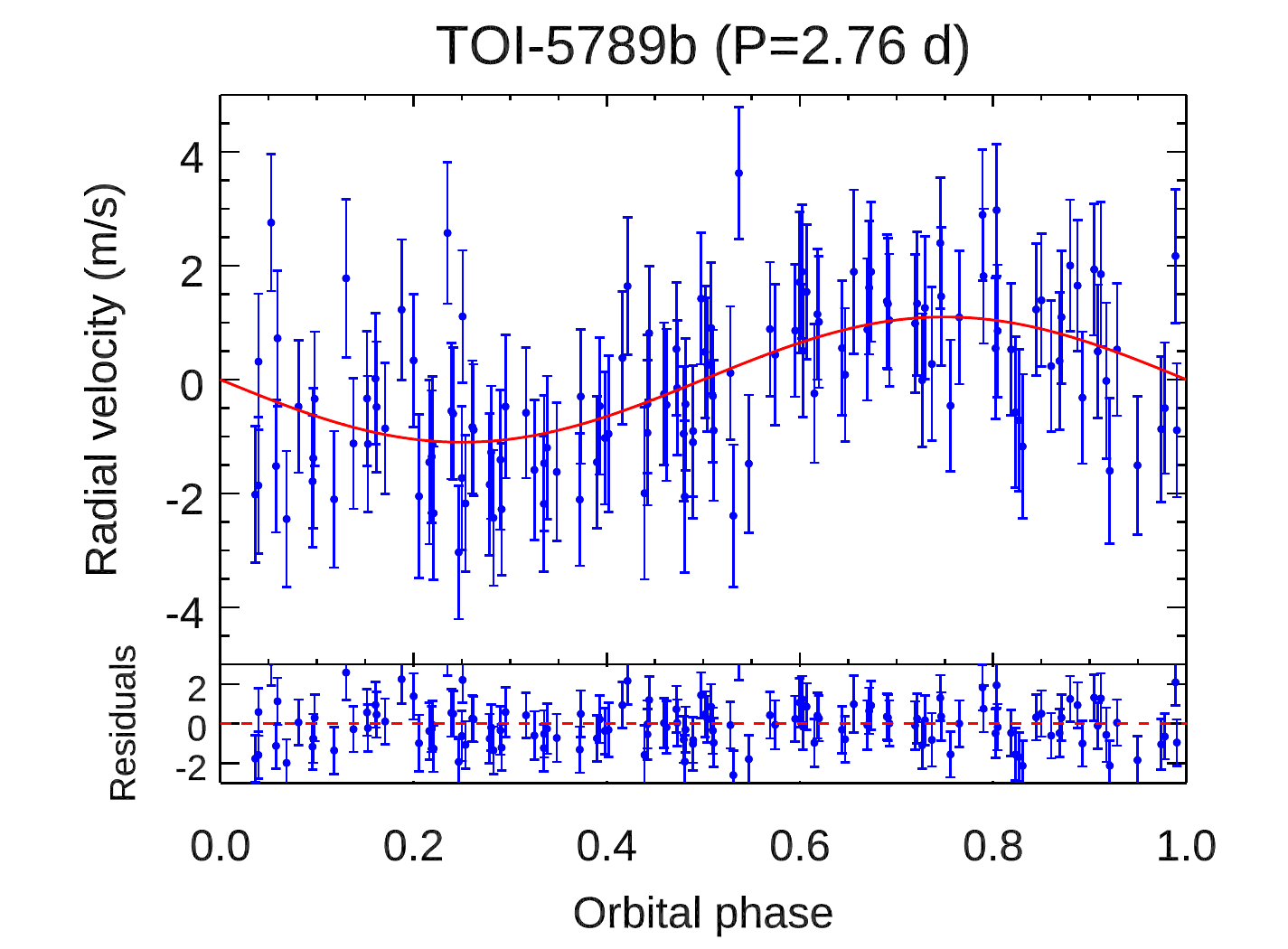}
\hspace{0.6cm}
\includegraphics[width=5.5 cm, angle=90]{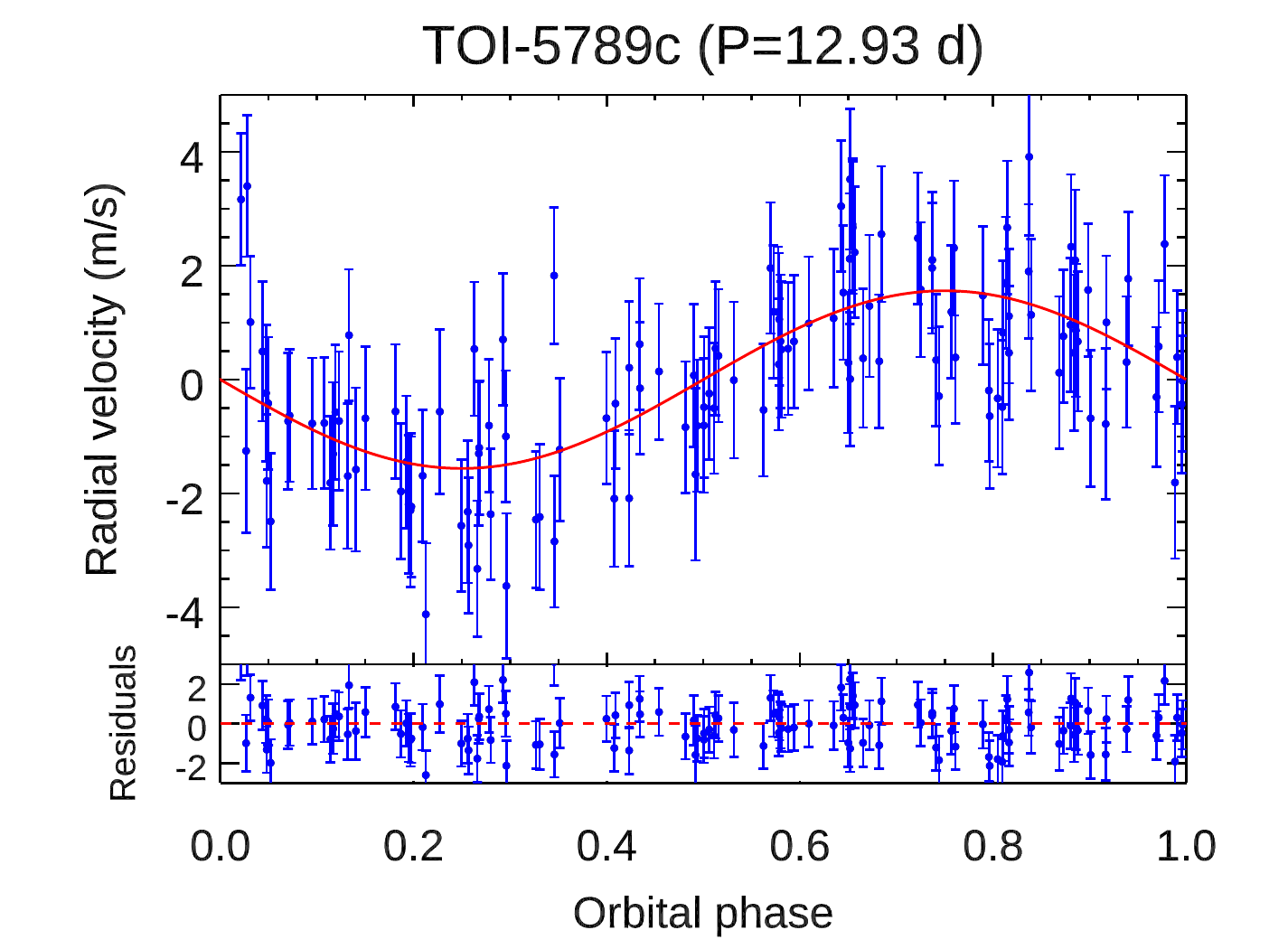} \\
\includegraphics[width=5.5 cm, angle=90]{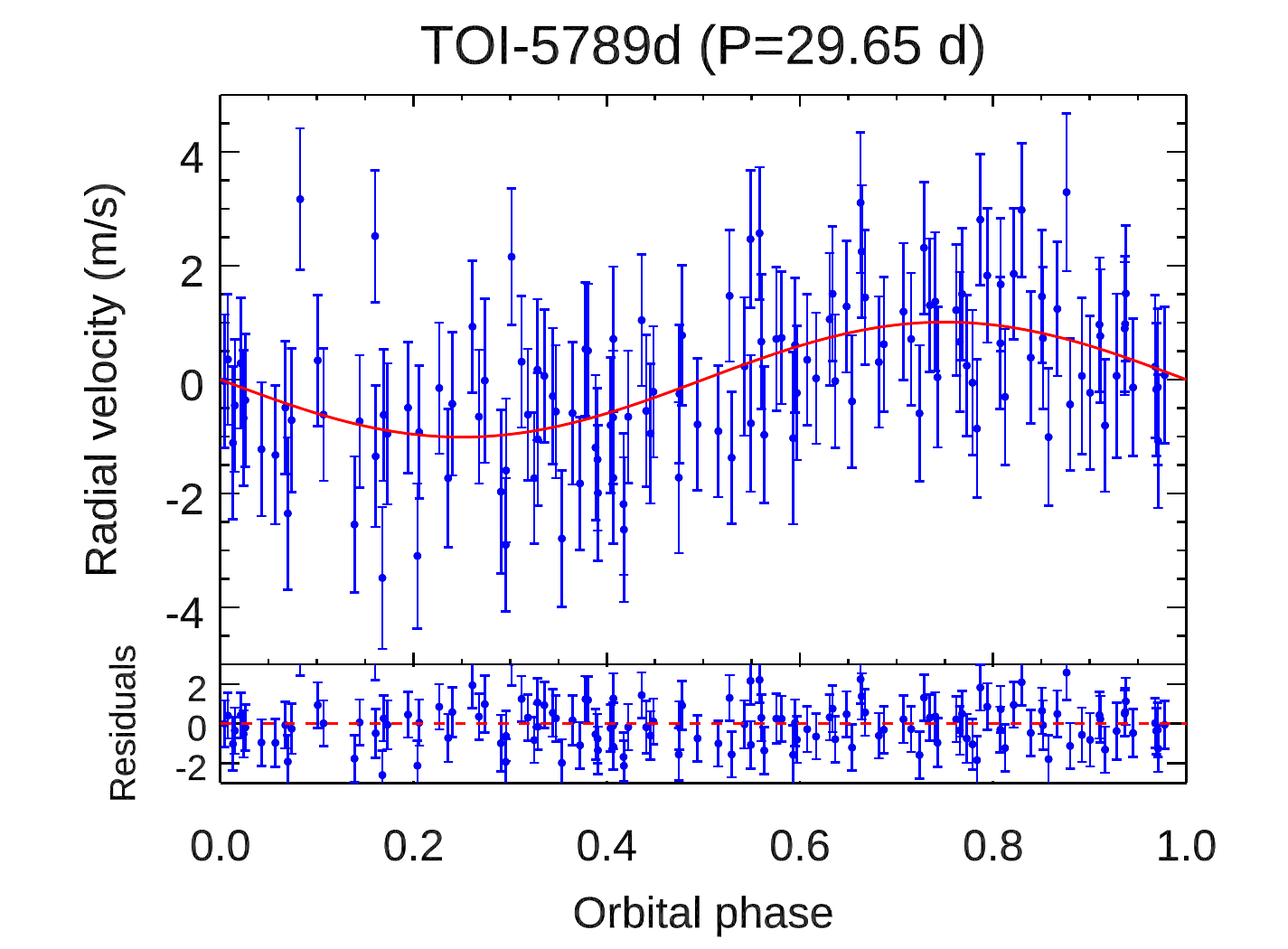}
\hspace{0.6cm}
\includegraphics[width=5.5 cm, angle=90]{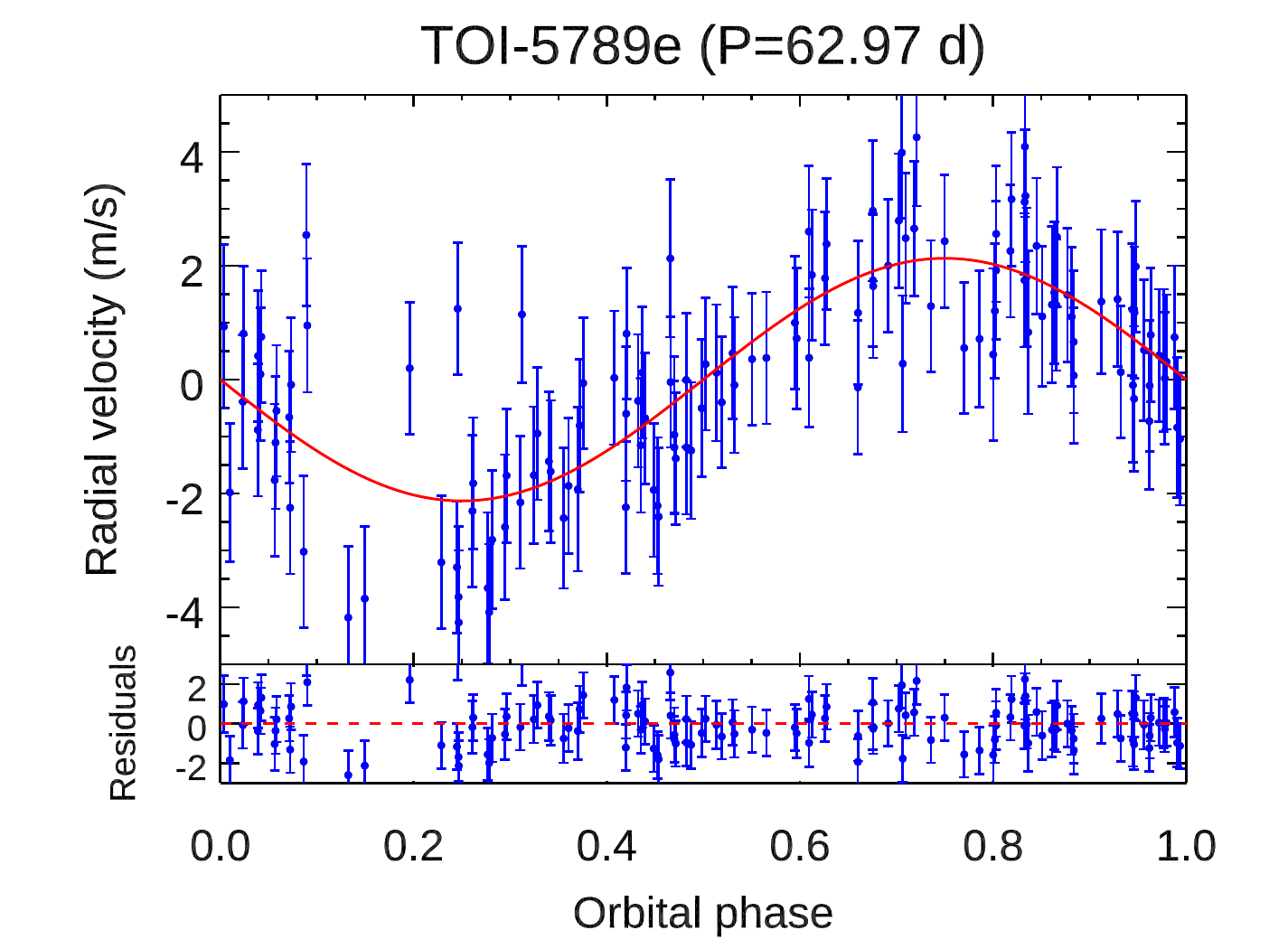}
\caption{HARPS-N radial-velocity signals of TOI-5789\,b (top left panel), c (top right panel),  d (bottom left panel), and e (bottom right panel) as a function of their orbital phase (phases 0 and 1 correspond to inferior conjunction times). The error bars take the radial-velocity jitter into account.}
\label{fig:RVs_phase}
\end{figure*}

The best values and uncertainties of the fit and derived parameters were determined as the medians and 15.86\%-84.14\% quantiles of their posterior distributions. They are given in Table~\ref{tab:planet_parameters} for the DE-MCMC analysis only, given that fully consistent parameters and uncertainties were found with \texttt{juliet}. The best model of the TOI-5789\,c transits is shown in Fig.~\ref{fig:transits_phase}; the best fit of the RVs as a function of both time and orbital phase with the four-planet model is displayed in Figs.~\ref{fig:HARPSN_RVs} and ~\ref{fig:RVs_phase}, respectively.

In summary, we find that the transiting sub-Neptune TOI-5789\,c has a radius of $R_{\rm c}=2.86^{+0.18}_{-0.15}~\rm R_\oplus$ and mass of $M_{\rm c}=5.00 \pm 0.50~\rm M_\oplus$ ($10\sigma$ precision), which correspond to a relatively low bulk density of $\rho_{\rm c}=1.16 \pm 0.23$~g\,cm$^{-3}$. The other non-transiting planets b, d, and e have minimum masses of $M_{\rm b}\sin{i}=2.12 \pm 0.28~\rm M_\oplus$ ($8\sigma$), $M_{\rm d}\sin{i}=4.29 \pm 0.68~\rm M_\oplus$ ($6\sigma$), and $M_{\rm e}\sin{i}=11.61 \pm 0.97~\rm M_\oplus$ ($12\sigma$). 

\begin{table}[ht!]
\tiny
\centering\renewcommand{\arraystretch}{1.3}
\caption{TOI-5789 system parameters.}     
\label{tab:planet_parameters}         
\renewcommand{\footnoterule}{}                          
\begin{tabular}{l c}        
\hline\hline                 
\emph{\textbf{Host star parameters}}  & Value and $1\sigma$ error \\
\hline
\hline
Systemic velocity $\gamma$ [\ms] &  $ -49315.60 \pm 0.11 $ \\
RV jitter $\sigma_{\rm RV, jit}$ [\ms] & $ 1.12 \pm 0.09$  \\
Phot. jitter $\sigma_{\rm phot, s54}$ & $7.11 \pm 1.28~\mbox{\sc{e}-05}$ \\
Phot. jitter $\sigma_{\rm phot, s81}$ & $1.50 \pm 0.18~\mbox{\sc{e}-04}$ \\
Limb-darkening coefficient $q_{1}$  & $0.40_{-0.16}^{+0.18}$ \\
Limb-darkening coefficient $q_{2}$  &  $0.51 \pm 0.35$  \\
Limb-darkening coefficient $u_{1}$  & $0.61_{-0.41}^{+0.44}$ \\
Limb-darkening coefficient $u_{2}$  &  $-0.01 \pm 0.42$  \\
\hline
\hline
\emph{\textbf{Planetary parameters}}  &  \\
\hline
\hline
TOI-5789\,b  & Value and $1\sigma$ error\\
\hline
Inferior conjunction time $T_{\rm c} \rm [BJD_{\rm TDB}-2450000]$ & $10358.054~(62)$ \\
Orbital period $P$ [days] & $2.76369~(96)$ \\
Orbital semimajor axis $a$ [au] & $0.03608 \pm 0.00045$  \\
Orbital eccentricity $e$  &  $\rm 0~(fixed)$   \\
RV semiamplitude $K$ [\ms] & $1.10 \pm 0.14$ \\
Planet minimum mass $M_{\rm p}\sin{i}~[\rm M_\oplus]$  &  $2.12\pm0.28$  \\
Equilibrium temperature$^{(1)}$ $T_{\rm eq}$ [K]  & $1201\pm26$  \\
\hline
TOI-5789\,c  &  \\
\hline
Transit Time $T_{\rm c} \rm [BJD_{\rm TDB}-2450000]$ & 10151.15868~(49) \\
Orbital period $P$ [days] & $12.927748~(16)$  \\
Transit duration $T_{\rm 14}$ [h] & $2.019 \pm 0.050$  \\
Radius ratio $R_{\rm p}/R_{*}$ & $0.0314_{-0.0013}^{+0.0018}$ \\
Inclination $i$ [deg] & $88.02 \pm 0.13$  \\
Impact parameter $b$ & $0.903 \pm 0.015$  \\
Orbital semimajor axis $a$ [au] & $0.1009 \pm 0.0013$  \\
Orbital eccentricity $e$  &  $ < 0.067$ \\
RV semiamplitude $K$ [\ms] & $1.56 \pm 0.15$ \\
Planet radius $R_{\rm p} ~[ \rm R_\oplus]$  &  $2.86_{-0.15}^{+0.18}$  \\
Planet mass $M_{\rm p} ~[\rm M_\oplus]$  &  $5.00 \pm 0.50$  \\
Planet density $\rho_{\rm p}$ [$\rm g\;cm^{-3}$] &  $1.16 \pm 0.23$  \\
Planet surface gravity log\,$g_{\rm p }$ [cgs] &  $2.77\pm0.06$  \\
Equilibrium temperature$^{(1)}$ $T_{\rm eq}$ [K]  & $718\pm15$  \\
\hline
TOI-5789\,d  &  \\
\hline
Inferior conjunction time $T_{\rm c} \rm [BJD_{\rm TDB}-2450000]$ & $10347.48 \pm 0.99$ \\
Orbital period $P$ [days] & $29.65\pm 0.12$ \\
Orbital semimajor axis $a$ [au] & $0.1756 \pm 0.0023$  \\
Orbital eccentricity $e$  &  $<0.096$  \\
RV semiamplitude $K$ [\ms] & $1.01 \pm 0.16$ \\
Planet minimum mass $M_{\rm p}\sin{i}~[\rm M_\oplus]$  &  $4.29\pm0.68$  \\
Equilibrium temperature$^{(1)}$ $T_{\rm eq}$ [K]  & $545\pm10$  \\
\hline
TOI-5789\,e  &  \\
\hline
Inferior conjunction time $T_{\rm c} \rm [BJD_{\rm TDB}-2450000]$ & $10359.05 \pm 1.17$ \\
Orbital period $P$ [days] & $62.98\pm0.22$  \\
Orbital semimajor axis $a$ [au] & $0.2901 \pm 0.0037$  \\
Orbital eccentricity $e$  &  $<0.071$  \\
RV semiamplitude $K$ [\ms] & $2.13 \pm 0.17$ \\
Planet minimum mass $M_{\rm p}\sin{i}~[\rm M_\oplus]$  &  $11.61\pm0.97$  \\
Equilibrium temperature$^{(1)}$ $T_{\rm eq}$ [K]  & $424\pm8$  \\
\hline
\hline
\end{tabular}
\begin{flushleft}
\footnotemark[1]{Blackbody equilibrium temperature assuming a null Bond albedo and uniform heat redistribution to the night side.} \\
\end{flushleft}
\end{table}

\section{Discussion}
\label{discussion}

\subsection{Architecture of the TOI-5789 planet system}
\label{architecture}
\begin{figure}[ht!]
    \centering
\includegraphics[width=0.48\textwidth]{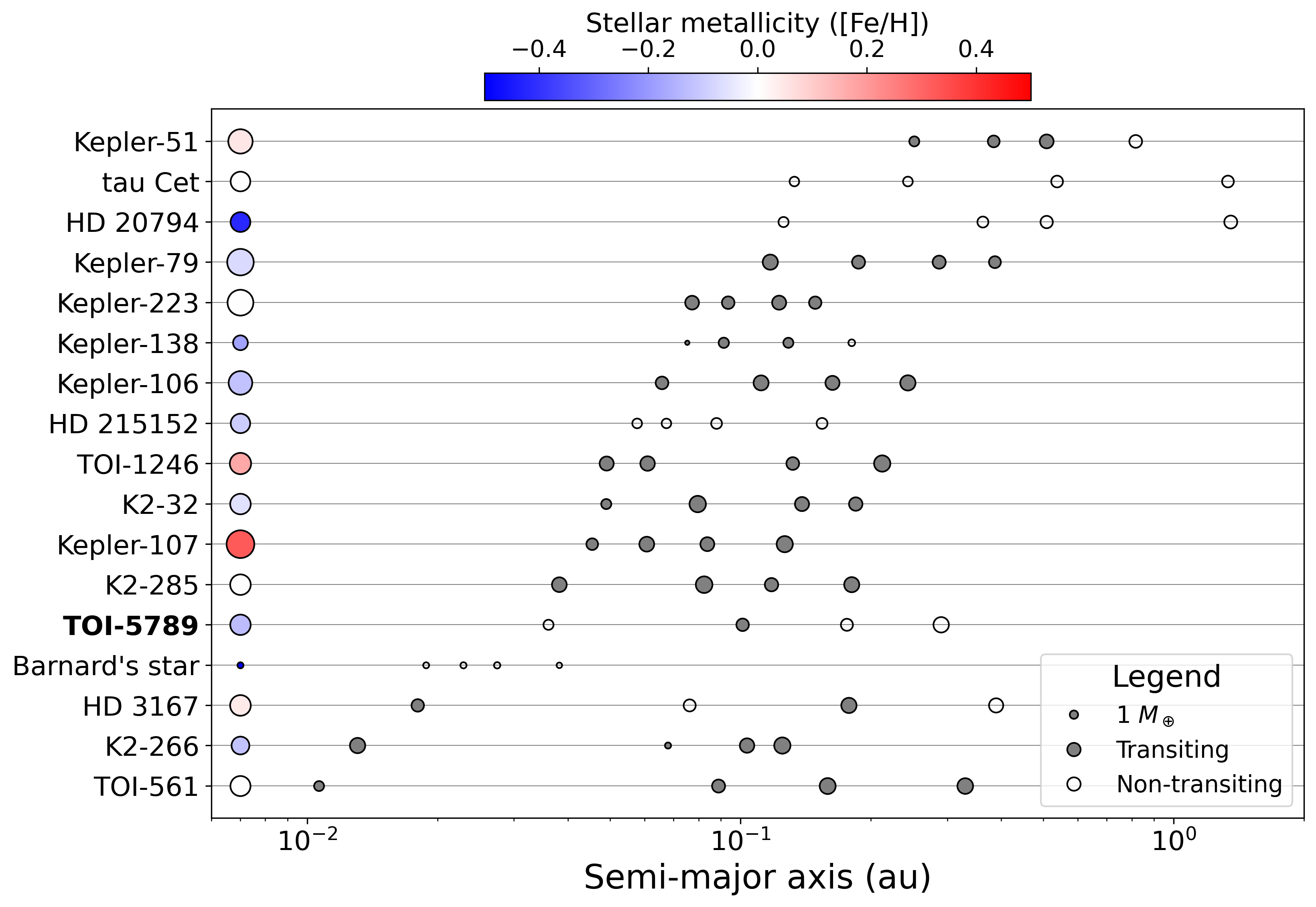}
    \caption{Architectures of known systems with four inner low-mass ($\mp < 20\,\Me$) planets. Filled and empty circles indicate transiting and non-transiting planets, respectively. The size of the circles is proportional to the planet mass or minimum mass (for the non-transiting planets). The circles representing the host stars on the left have sizes proportional to the stellar mass, and are color coded as a function of stellar metallicity (metallicity
increases from blue to red).}
    \label{fig:arch_4pl_sys}
\end{figure}

By comparing the TOI-5789 system with the other known systems hosting four inner low-mass ($\mp < 20\,\Me$) planets in Fig.~\ref{fig:arch_4pl_sys}, it can be immediately noted that TOI-5789 is the only system with at least one transiting planet, where the innermost planet (TOI-5789\,b) is not transiting. Since the inclination of planet b must be $i_{\rm b} < 83.8$\,deg (corresponding to $b_{\rm b}>1$), its mutual inclination with planet c must be higher than 4\,deg. In contrast, the typical mutual inclinations of the \emph{Kepler} systems with two or more transiting planets have a median of $\lesssim 2$\,deg (e.g., \citealt{Fabrycky2014, He2021}). This suggests that TOI-5789 system might be the remnant of a metastable tightly packed inner system that underwent a phase of dynamical instability, leading to its consolidation into a stable configuration after the ejection or collision of one or more planets \citep{Volk2015, Pu2015, RiceSteffen2023}.

\begin{figure}[ht!]
    \centering
\includegraphics[width=0.5\textwidth]{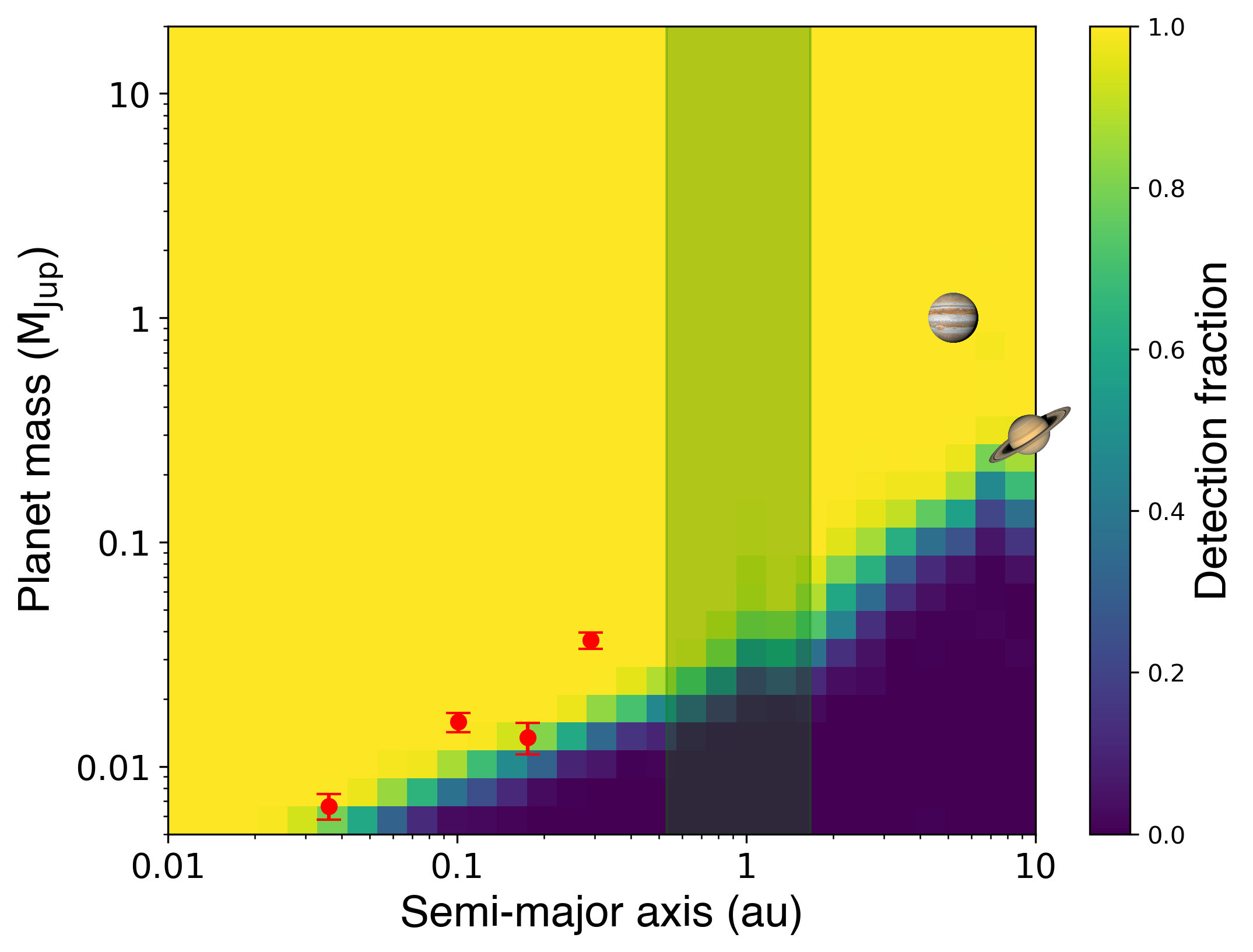}
\caption{Completeness map of TOI-5789, showing the sensitivity of the HARPS-N and HIRES radial-velocity measurements to the presence of planets as a function of their mass and orbital separation, through experiments of injection and recovery of Doppler signals. The detectability inside the cell is indicated by its color, according to the colorbar on the right-hand side of the figure. The four red circles indicate the detected planets TOI-5789\,b, c, d, and e, while the green shade represents the optimistic habitable zone.}
    \label{fig:completeness}
\end{figure}

The high mutual inclinations do not seem to be caused by an undetected outer gaseous giant planet ($\mp \geq 0.3$\,\Mjup), whose presence can be excluded by the RV sensitivity map in Fig.~\ref{fig:completeness}. The latter was computed with both HARPS-N and HIRES data (for their longer timespan), following \citet{Bryan2019} and \citet{Bonomo2023}. 
The lack of a statistically significant Hipparcos-Gaia DR3 astrometric acceleration \citep{Brandt2021,Kervella2022} also allows to rule out the presence of $>0.1-0.2$\,\Mjup\, companions in the $3-10$\,au range around TOI-5789. 
After all, cold giant planets in low-mass planet systems are rare around host stars with sub-solar mass and metallicity, such as TOI-5789 \citep{BryanLee2025, Bonomo2025b}.
From the same sensitivity map we can also exclude the presence of planets with $\mp \sim 12\,\Me$ in the optimistic habitable zone \citep{Kopparapu2013}.

\subsection{A possible formation history of TOI-5789}
\label{formation}
We used our Monte Carlo version of the \texttt{GroMiT} (planetary Growth and Migration Tracks) code \citep{Polychroni2023} to simulate the possible tracks of the four TOI-5789 planets whilst forming in the framework of the pebble accretion scenario (see, e.g.,  \citealt{Damasso2024, Naponiello2025}). We used the treatments of \citet{Johansen2019} and \cite{Tanaka2020} for the growth and migration of pebble and gas accreting planets as well as the scaling law for the pebble isolation mass from \citet{Bitsch2018}. The model uses a viscously evolving protoplanetary disk \citep{Johansen2019, Armitage2020} with prescriptions for  its solid-to-gas ratio profile \citep{Turrini2023} and its thermal profile due to the interplay between viscous heating and stellar irradiation \citep{Ida2018}.

We ran a set of 10$^5$ individual planets varying the initial time at which the planetary seed of 0.01\,M$_\oplus$ was inserted in the disk ($10^4 - 3\cdot10^6$\,yrs), the initial semimajor axis of the seed (0.1--40\,au), the disk viscosity ($\alpha=3\cdot10^{-3} - 3\cdot10^{-4}$), and the pebble size (sub-mm to cm). We set i) the mass of the protoplanetary disk to 5\% of the stellar mass; ii) the stellar luminosity during the pre-main sequence to 1.014\,$L_{\odot}$, based on the evolutionary models from \citet{Baraffe2015} for stars with the mass of TOI-5789 at 1.5\,Myr; iii) the disk characteristic radius, R$_0$, to 40\,au, 
and the disk temperature at 1\,au, T$_0$, to  160\,K, following  \citet{Johansen2019}; and iv) the exponents of both the surface density profile and the disk temperature T$_0$ to 0.8 and 0.5, respectively  \citep{Turrini2019}. 

We find that the outermost planet (planet e) formed both earlier and at a larger orbital distance than the innermost one (planet b). In particular, we find that planet e most likely formed beyond the snowline, and planet b within it. The intermediate planets c and d could have formed either inside or outside the water snowline, and as such, spectral characterization is needed to disentangle this degeneracy (see Sect.~\ref{atmospheric_characterization}). The simulations tend to favor disks with viscosity larger than $10^{-3}$ and  pebbles with predominantly mm or cm sizes, in order to form all four planets.

\subsection{Composition of TOI-5789\,c}
\begin{figure}[h!]
\hspace{-0.5 cm}
\includegraphics[width=8.0cm, angle=90]{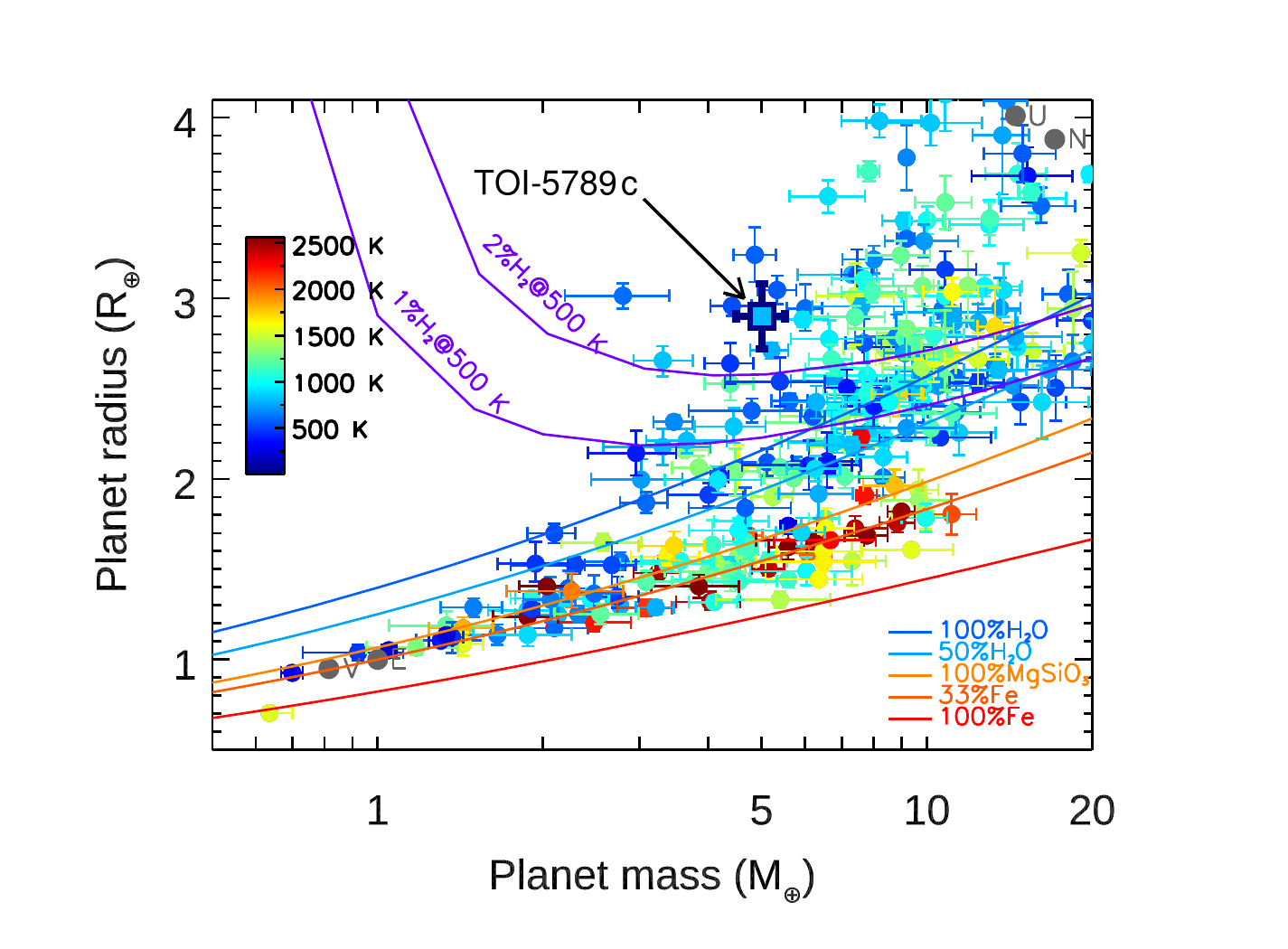}
\vspace{-0.5cm}
\caption{Mass-radius diagram of small ($R_{\rm p} \leq 4~\Re$) planets with mass and radius determinations better than $4\sigma$ and $10\sigma$, respectively, color-coded by planet equilibrium temperatures. 
The different solid curves, from bottom to top, correspond to planet compositions of $100\%$ iron, $33\%$ iron core and $67\%$ silicate mantle (Earth-like composition), $100\%$ silicates, $50\%$ rocky interior and $50\%$ water, $100\%$ water, Earth-like rocky interiors and $1\%$ or $2\%$ hydrogen-dominated atmospheres \citep{Zeng2013}.
The gray dark circles indicate Venus (V), the Earth (E), Uranus (U), and Neptune (N). 
TOI-5789\,c is indicated with a square.}
\label{fig:massradius_diagram}
\end{figure}

Figure~\ref{fig:massradius_diagram} shows the mass-radius diagram of small exoplanets with mass and radius determinations better than $4\sigma$ and $10\sigma$, respectively, along with the theoretical mass-radius curves  from \citet{Zeng2013}. 
TOI-5789\,c is located above the isocomposition curve of an Earth-like rocky
interior with a 2\% H$_2$-dominated atmosphere, consistent with its relatively low mean density (Table~\ref{tab:planet_parameters}).

We analyzed the potential composition of TOI-5789\,c in more detail using an inference model based on \citet{Dorn2017} with updates described in \citet{luo_majority_2024}.
The underlying forward model consists of three layers: an iron core, a silicate mantle, and a H$_2$-He-H$_2$O atmosphere.
For the solid phase of the iron core, we used the equation of state (EOS) of hexagonal close packed iron \citep{hakim_new_2018,miozzi_new_2020}. 
For the liquid iron phase, we used the EOS from \citet{luo_majority_2024}. 
The silicate mantle is composed of three major species: MgO, SiO$_2$, and FeO. 
We modeled the solid phase of the mantle using the thermodynamical model \textsc{Perple\_X} \citep{connolly_geodynamic_2009} for pressures below $\approx 125\,$GPa, while for higher pressures we defined the stable minerals a priori and used their respective EOS from various sources \citep{hemley_constraints_1992,fischer_equation_2011,faik_equation_2018,musella_physical_2019}.
The liquid mantle was modeled as a mixture of Mg$_2$SiO$_4$, SiO$_2$, and FeO, as there is no data for the density of liquid MgO in the required pressure-temperature regime \citep{melosh_hydrocode_2007,faik_equation_2018,ichikawa_ab_2020,stewart_shock_2020}. 
In all cases, the EOS of the different components were mixed using the additive volume law for ideal gases. 
Both the iron core and the silicate mantle were modeled as adiabatic. For this application, we fixed an Earth-like core-to-mantle ratio of 0.325:0.675.

The H$_2$-He-H$_2$O atmosphere layer was modeled using the analytic description by \citet{guillot_radiative_2010} and \citet{2014_Jin_planetarypopulation}, and consists of an irradiated layer on top of a non-irradiated layer in radiative-convective equilibrium. 
The water mass fraction is given by $Z$, and the hydrogen-helium ratio is set to solar. 
The two components of the atmosphere, H$_2$/He and H$_2$O, were again mixed following the additive volume law. 
We used the EOS by \citet{1995_Saumon_EOS} for H$_2$/He and the ANalytic Equations of States (ANEOS, cf.  \citealt{1990_thompson_aneos}) for H$_2$O.

For the inference, we used  Polynomial Chaos Kriging surrogate modeling to approximate the global behavior of the full physical forward model and replaced it in the MCMC framework \citep{DeWringer}. 
The prior parameter distribution is listed in Table~\ref{tab:comppriors}, and the results of the inference model are summarized in Table~\ref{tab:inferenceresult} and Figure \ref{fig:composition}.
As expected from the position of the planet in the mass-radius diagram (Fig.~\ref{fig:massradius_diagram}), we find that TOI-5789\,c possesses a significant atmosphere, even though the exact mass is relatively unconstrained with $M_\mathrm{atm} = $0.19$^{+0.16}_{-0.11} \,M_\oplus$, which represents a mass fraction of $\sim4.6\%$.
The composition of this atmosphere is likely super-solar fitting a mass fraction of H$_2$O of $Z = 0.26 ^{+0.26}_{-0.17}$.

\begin{table}
\centering
\caption{Inference results for the internal composition of TOI-5789\,c with $1\sigma$ uncertainties.} 
\label{tab:inferenceresult}
\renewcommand{\arraystretch}{1.2}
\begin{tabular}{lc}
    \hline\hline
     Parameters & TOI-5789\,c  \\
    \hline \\[-6pt]%
     log($M_\mathrm{atm}/ M_\oplus$) & $-0.73^{+0.28}_{-0.34} $ \\
    $M_\mathrm{atm} /M_\oplus$ & $0.19^{+0.16}_{-0.11} $ \\
    $M_\mathrm{core+mantle}/ M_\oplus$ & $4.85^{+0.48}_{-0.5}$ \\
    $Z_\mathrm{env}$ (envelope water mass fraction) & $0.26 ^{+0.26}_{-0.17} $ \\
    \bottomrule
\end{tabular}
\end{table}

\subsection{Prospects for atmospheric characterization of TOI-5789c}
\label{atmospheric_characterization}
\begin{figure}[ht!]
    \centering
\includegraphics[width=0.5\textwidth]{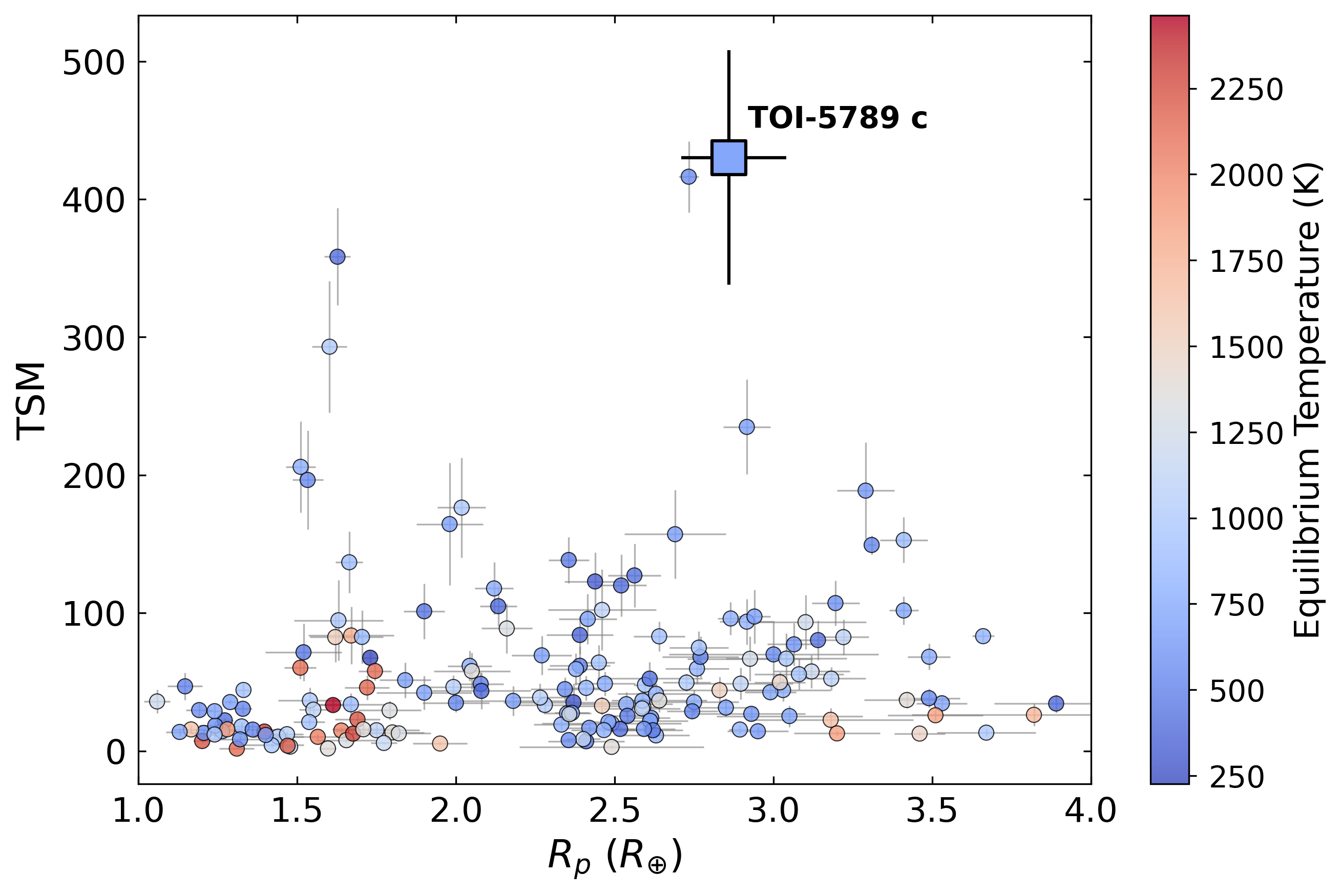}
    \caption{Transmission spectroscopy metric (TSM) of small planets with $R_{\rm p} \leq 4 \rm R_{\oplus}$ (circles). TOI-5789\,c is shown with a large square. The colors of the symbols indicate planet equilibrium temperatures, according to the color bar on the right.}
    \label{fig:TSM}
\end{figure}

The combination of favorable $\rho_{\rm p}$ and $T_\mathrm{eq}$ parameters with the stellar brightness in the $K_s$ band (Table~\ref{tab:star}) makes TOI-5789\,c a prime candidate for atmospheric characterization. It currently has the highest transmission spectroscopy metric (TSM=$430^{+92}_{-78}$, \citealt{KemptonEtal2018paspTransitSpectroscopicMetric}) among the small planets with $R_{\rm p} \leq 4 \rm R_{\oplus}$ (see Fig.~\ref{fig:TSM}), the closest planet being GJ\,1214b (e.g., \citealt{Ohno2025}). 

Here we explore the prospect for future characterization with the James Webb Space Telescope (JWST, \citealt{Gardner2006}) and the Ariel mission \citep{Tinetti2018}. 
We performed atmospheric retrievals of simulated transmission spectra at 1$\times$ and 100$\times$
atmospheric solar metallicity to show the difference between the two scenarios, and to assess the confidence at which we can constrain the atmospheric composition. 

The model atmospheres consist of a 1D set of cloud-free isothermal
layers in hydrostatic equilibrium, with constant volume-mixing ratio
(VMR) profiles.  We included molecular opacities for \ch{H2O}
\citep{PolyanskyEtal2018mnrasPOKAZATELexomolH2O}, \ch{CO}
\citep{LiEtal2015apjsCOlineList}, \ch{CO2}
\citep{YurchenkoEtal2020mnrasCO2ucl4000}, \ch{CH4}
\citep{YurchenkoEtal2024mnrasExomolCH4mm}, and \ch{SO2}
\citep{UnderwoodEtal2016mnrasSO2exoamesExomol}; as well as Rayleigh
\citep{Kurucz1970saorsAtlas} and collision-induced absorption
\citep{BorysowEtal2001jqsrtH2H2highT, Borysow2002jqsrtH2H2lowT,
  RichardEtal2012jqsrtCIA}.

\subsubsection{JWST simulations}
For the JWST simulations, we employed the \textsc{Pyrat Bay} framework
to simulate the transmission spectra
\citep{CubillosBlecic2021mnrasPyratBay}.  We then used the Gen~TSO
tool \citep{Cubillos2024paspGenTSO} to estimate the noise of the
spectra as observed by JWST.  We identified the NIRCam's Grism Time
Series mode as an optimal configuration to use, given the possibility
to observe simultaneously at short and long wavelengths, providing
nearly continuous 1.0--5.0\,{\micron} coverage with two transit
observations: a combination of the F150W2+F322W2 and F150W2+F444W
filters (Fig.\,\ref{fig:atmospheric_retrievals}, left panel).

We then performed atmospheric retrievals of the simulated JWST spectra with Pyrat Bay employing the Multinest Nested Sampling algorithm \citep{Feroz2009MNRAS.398.1601F}.  The retrieval free parameters consisted of the isothermal temperature, the reference pressure level at $R_{\rm p}$, and the volume mixing ratios (VMRs) of the five molecules listed above.  We find that the combination of these three NIRCam filters is sensitive to the abundance of all \ch{H2O}, \ch{CO2}, CO, \ch{CH4}, and \ch{SO2} species.  For both metallicity scenarios, we are able to constrain the molecular abundances with a precision of 0.1--0.3 dex (Fig.\,\ref{fig:atmospheric_retrievals}, right panel).

\subsubsection{Ariel simulations}
TOI-5789\,c is also part of the current Ariel mission Candidate Reference Sample (CRS), and so we assessed Ariel’s capability to constrain the chemistry of its atmosphere. For the Ariel simulations, we used the open-source code TauREx \citep{Al-Refaie2021ApJ...917...37A,Al-Refaie2022ApJ...932..123A} to generate forward spectra, and then performed inverse spectral retrievals for the same 1$\times$ and 100$\times$ solar-metallicity scenarios as above. First, we produced a high-resolution ($R = 15000$) forward model. We then convolved this spectrum with normally distributed noise simulations generated by the official Ariel noise simulator ArielRad (ArielRad v. 2.4.26, \citealt{Mugnai2020ExA....50..303M}, Ariel Payload v. 0.0.17, ExoRad v 2.1.111, \citealt{Mugnai2023JOSS....8.5348M}) to obtain simulated observed data at the instrumental resolution.
In this study, we estimated the spectra at Ariel's Tier-3 resolution \citep{Tinetti2022EPSC...16.1114T}. At this tier, the raw spectral data are binned to $R = 20$, 100, and 30 in FGS-NIRSpec (1.1-1.95\,\micron), AIRS-Ch0 (1.95-3.9\,\micron), and AIRS-Ch1 (3.9-7.8\,\micron), respectively. By using ArielRad we find that five transit observations are required to achieve the Tier-3 S/N for TOI-5789\,c.
We performed atmospheric retrievals with TauREx in retrieval mode, using MultiNest \citep{Feroz2009MNRAS.398.1601F}, and broad uninformative priors to recover the atmospheric composition. 
The retrievals indicate that in the 1$\times$ solar scenario, the abundances of all molecules can be effectively constrained, with precision and accuracy comparable to JWST (Fig.~\ref{fig:ariel_atmospheric_retrievals}). In the 100$\times$ solar scenario, however, the results of our retrievals are less accurate, and constraining the CO mixing ratio is more challenging (Fig.~\ref{fig:ariel_atmospheric_retrievals}). 
This is likely due to the Ariel lower resolution beyond $\sim 4$\,\micron, which makes it more difficult to separate the contributions of different molecules. In particular, we note that the underestimation of the CO abundance is related to a slight overestimation of the abundances of H$_2$O and CO$_2$, whose spectral features partially overlap with the CO absorption  near 4.5\,{\micron} (see Fig.~\ref{fig:ariel_atmospheric_retrievals}).
~\\

JWST NIRCam and Ariel observations of TOI-5789\,c would be highly complementary. On the one hand, JWST spectra will have a considerably higher S/N than Ariel because of the JWST larger aperture. On the other hand, Ariel will gather spectrophotometry in a wider spectral range, from $\sim0.6$ to $\sim8$\,{\micron}, in a single observation. The range from $\sim 5$ to 8\,{\micron} is sensitive to additional bands of H$_2$O, CH$_4$, and especially SO$_2$, despite the Ariel lower resolution mentioned above. Furthermore, the three Ariel bands below $\sim 1$\,{\micron} may help to better characterize the possible presence of hazes and/or clouds. Combined retrievals of both JWST NIRCam and Ariel transmission spectra will certainly improve the accuracy and precision of molecular abundances, and hence of the planet atmospheric metallicity (a detailed investigation of combined retrievals goes, however, beyond the scope of this work).

\section{Summary and conclusions}
\label{conclusions}
In this work, we confirm the planetary nature of the TESS sub-Neptune ($R_{\rm p}\simeq2.9~\rm R_\oplus$) candidate TOI-5789.01, now designated as TOI-5789\,c, which transits the bright and old K1\,V star HIP\,99452  every 12.93\,d. With 141 HARPS-N RVs we determine its mass and bulk density to be $M_{\rm p}=5.00 \pm 0.50~\rm M_\oplus$ and $\rho_{\rm p}=1.16 \pm 0.23$~g\,cm$^{-3}$, respectively. From the fundamental parameters of the planet we infer that TOI-5789\,c must possess a significant atmosphere, although the exact atmospheric mass fraction is not well constrained.

Thanks to the stellar brightness in the near-IR and relatively low planet density, we show that TOI-5789\,c is a favorable target for atmospheric characterization with both JWST and, in the future, Ariel transmission spectroscopy, assuming a cloud-free atmosphere. 
Atmospheric characterization of sub-Neptunes is crucial to break degeneracies in their bulk compositions, allowing us to distinguish between gas-rich dwarfs and water-rich worlds.

Last but not least, thanks to numerous and diverse analyses of the HARPS-N RVs, we show that the TOI-5789 system contains three more non-transiting planets. We note that the mutual orbital inclination between the innermost planet TOI-5789\,b and TOI-5789\,c is higher than the typical mutual inclinations in \emph{Kepler} compact multiplanet systems. 
The relatively high mutual inclinations in the TOI-5789 system should be unrelated to any outer gas giants.

The detection of low-mass planets around bright stars is one of the main goals of the PLATO space mission \citep{Rauer2014, Rauer2025}. The expected growing number of sub-Neptunes around bright hosts, which are thus suited to both high-precision RV follow-up and detailed atmospheric characterization, will enable us to gain an ever-better understanding of the nature and formation of the sub-Neptune planets.


\section*{Data availability}
Table~\ref{tab:data} is only available in electronic form at the CDS via anonymous ftp to \url{cdsarc.u-strasbg.fr} (130.79.128.5) or via \url{http://cdsweb.u-strasbg.fr/cgi-bin/qcat?J/A+A/}.


\begin{acknowledgements}
This work is based on observations made with the Italian Telescopio Nazionale Galileo (TNG) operated on the island of La Palma by the Fundaci\'{o}n Galileo Galilei of the INAF at the Spanish Observatorio del Roque de los Muchachos of the Instituto de Astrofisica de Canarias (GTO programme). 
The HARPS-N project was funded by the Prodex Program of the Swiss Space Office (SSO), the Harvard-University Origin of Life Initiative (HUOLI), the Scottish Universities Physics Alliance (SUPA), the University of Geneva, the Smithsonian Astrophysical Observatory (SAO), the Italian National Astrophysical Institute (INAF), the University of St. Andrews, Queen's University Belfast and the University of Edinburgh.
This paper is based on data collected by the TESS mission. Funding for the TESS mission is provided by the NASA Science Mission directorate. We acknowledge the Italian center for Astronomical Archives (IA2, https://www.ia2.inaf.it), part of the Italian National Institute for Astrophysics (INAF), for providing technical assistance, services and supporting activities of the GAPS collaboration.
This paper makes use of observations made with the Las Cumbres Observatory’s education network telescopes that were upgraded through generous support from the Gordon and Betty Moore Foundation.
This research has made use of the Exoplanet Follow-up Observation Program (ExoFOP; DOI: 10.26134/ExoFOP5) website, which is operated by the California Institute of Technology, under contract with the National Aeronautics and Space Administration under the Exoplanet Exploration Program.
The authors acknowledge financial contribution from the INAF Large Grant 2023 ``EXODEMO'' and the INAF Guest Observer Grant (Normal) 2024 ``ArMS: the Ariel Masses Survey Large Program at the TNG''. 
A.S.B. and L.Man. acknowledge financial contribution from the European Union - Next Generation EU RRF M4C2 1.1 PRIN MUR 2022 projects 2022CERJ49 (ESPLORA) and 2022J4H55R, respectively.
C.D.M. acknowledges support from the INAF Mini-Grant ``Impact of planetary Masses and Radii Estimates on the Atmospheric Retrievals – IMaREA'', awarded under the INAF Fundamental Astrophysics funding scheme. Part of the research activities presented in this paper were carried out with the contribution of the Next Generation EU funds within the Italian National Recovery and Resilience Plan (PNRR), Mission 4 – Education and Research, Component 2 – From Research to Business (M4C2), Investment Line 3.1 – Strengthening and Creation of Research Infrastructures, Project IR0000034 ``STILES – Strengthening the Italian Leadership in ELT and SKA''.
K.A.C. acknowledges support from the TESS mission via subaward s3449 from MIT. 
E.P. acknowledges financial support from the Agencia Estatal de Investigaci\'on of the Ministerio de Ciencia e Innovaci\'on MCIN/AEI/10.13039/501100011033 and the ERDF ``A way of making Europe” through project PID2021-125627OB-C32, and from the Centre of Excellence ``Severo Ochoa” award to the Instituto de Astrofisica de Canarias. T.Z. acknowledges support from CHEOPS ASI-INAF agreement n. 2019-29-HH.0, NVIDIA Academic Hardware Grant Program for the use of the Titan V GPU card, and the Italian MUR Departments of Excellence grant 2023-2027 ``Quantum Frontiers''. The authors wish to thank the anonymous referee for her/his comments, which allowed them to improve their manuscript;  Dr.~A.~Bocchieri for comments to the paper; and Dr.~S.~Messina for discussions about  the stellar magnetic activity.
\end{acknowledgements}

\bibliographystyle{aa} 
\bibliography{aa57662-25} 

\clearpage

\appendix

\onecolumn

\section{Data}

\begin{figure*}[ht!]
\centering
\vspace{1cm}
\includegraphics[width=16.25 cm]{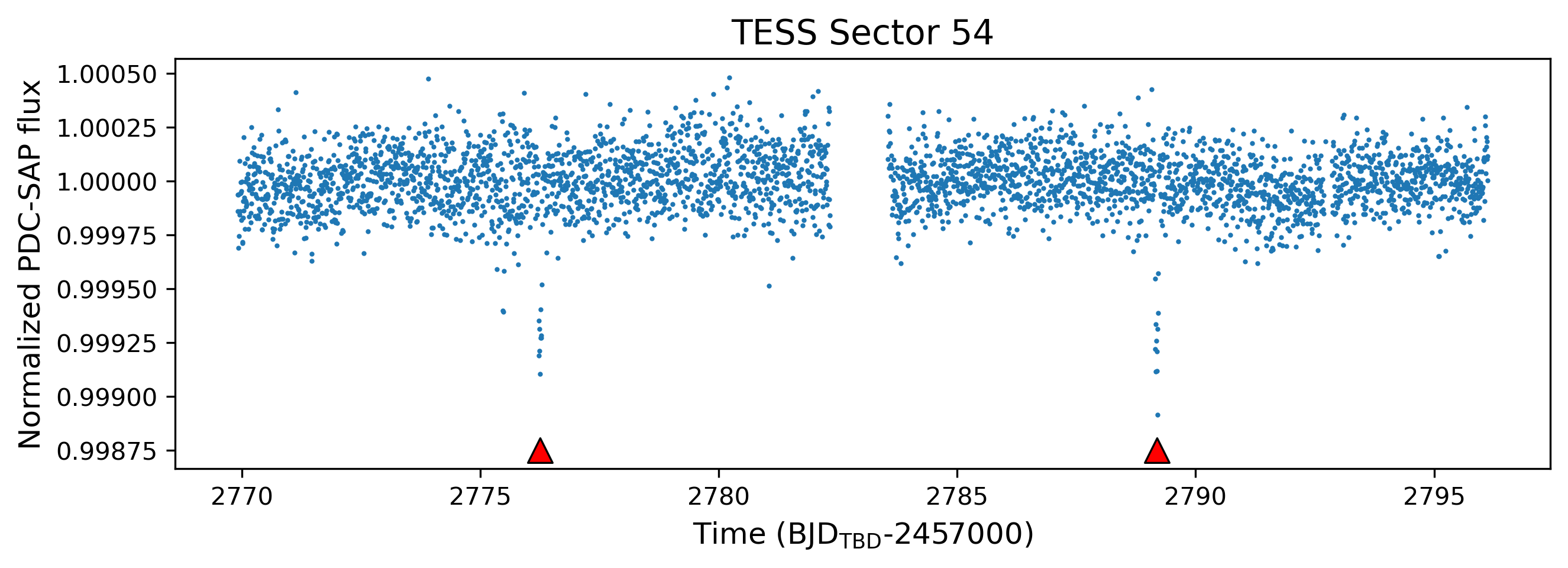}
\caption{PDC-SAP TESS light curve from Sector 54. The transits of TOI-5789\,c are marked with red arrows.}
\label{fig:sector_54}
\end{figure*}

\begin{figure*}[ht!]
\centering
\includegraphics[width=16 cm]{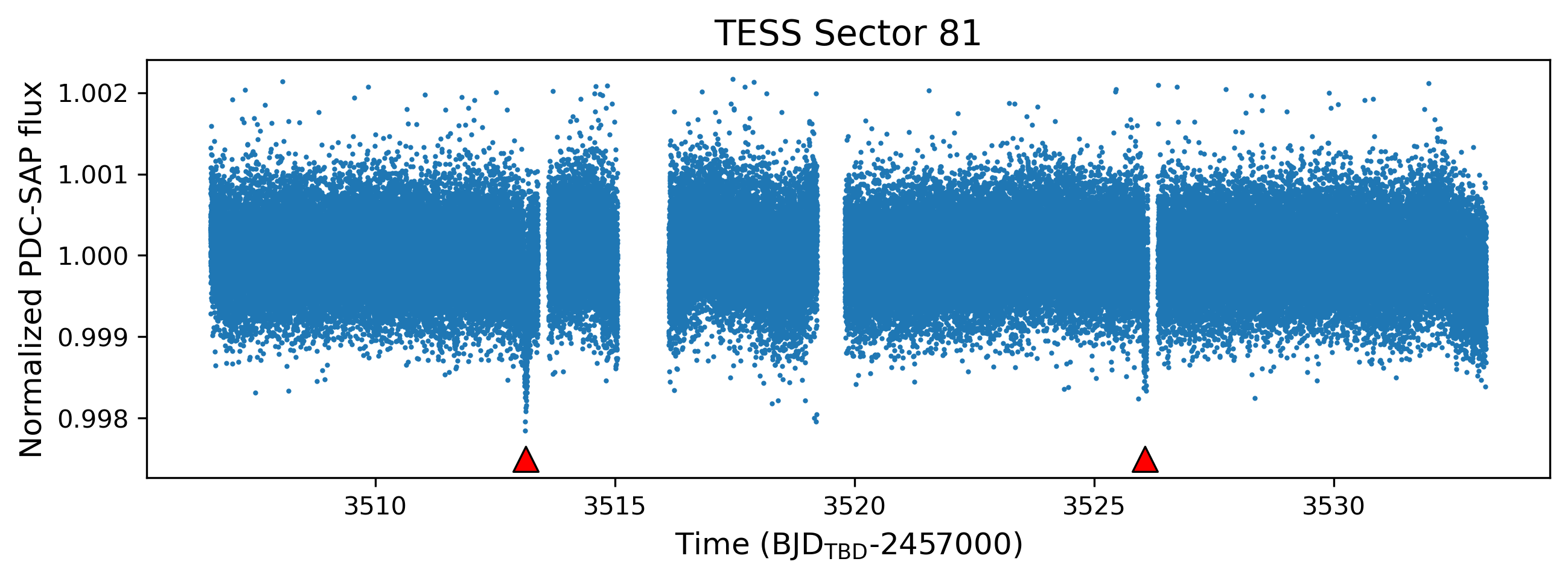}
\caption{PDC-SAP TESS light curve from Sector 81. The transits of TOI-5789\,c are marked with red arrows.}
\label{fig:sector_81}
\end{figure*}

\begin{figure*}[h!]
\centering
\includegraphics[width=8.0cm]{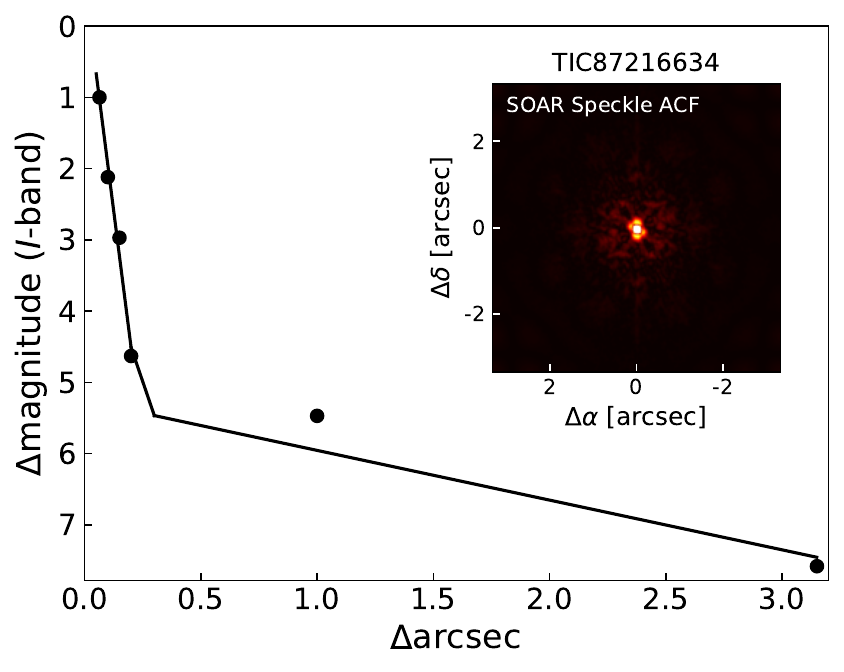}
\hspace{+0.25 cm}
\includegraphics[width=8.75cm]{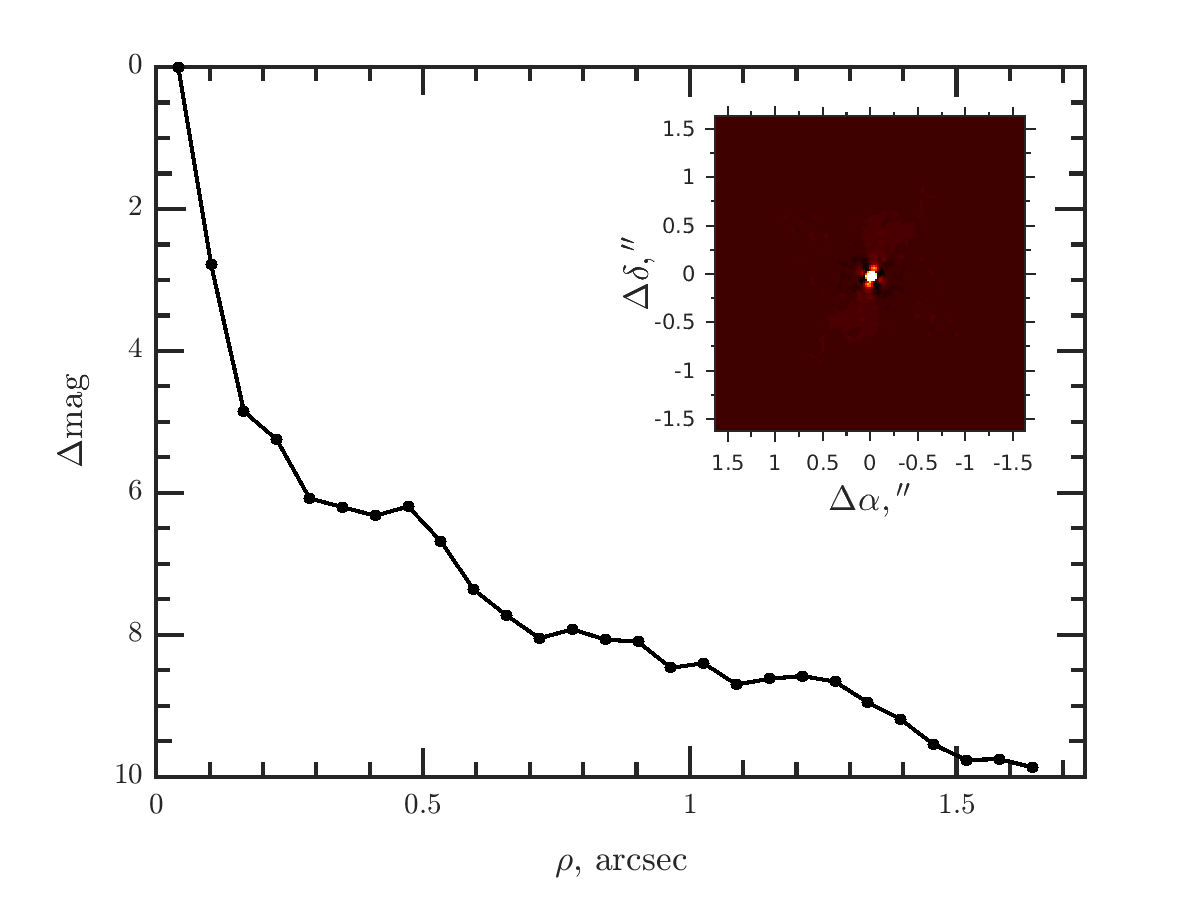}
\caption{Detection sensitivity from the SOAR (left) and SAI (right) high-angular resolution images.}
\label{fig:det_sensitivity_imaging}
\end{figure*}

\clearpage

\onecolumn

\begin{figure*}
\begin{multicols}{2}
\centering
\includegraphics[width=0.40\textwidth, angle=90]{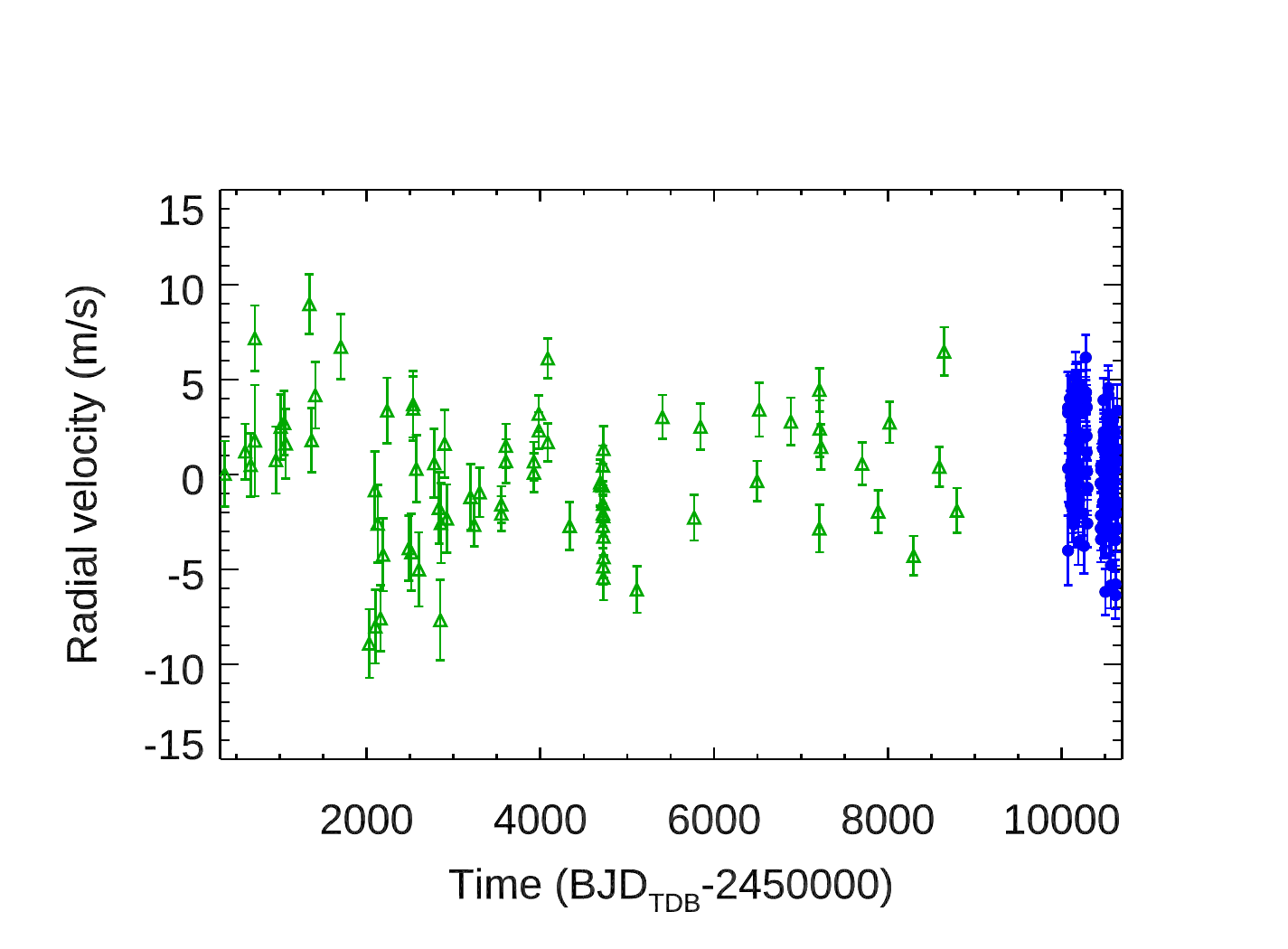}\par
\caption{Radial velocities of TOI-5789 gathered with the HIRES (green triangles) and HARPS-N (blue circles) spectrographs. Radial-velocity zero points as determined with the DE-MCMC analysis (Sect.~\ref{DE-MCMC_analysis}) were subtracted from each dataset. }
\label{fig:HARPSN_HIRES_RVs}
\centering
\includegraphics[width=0.55\textwidth]{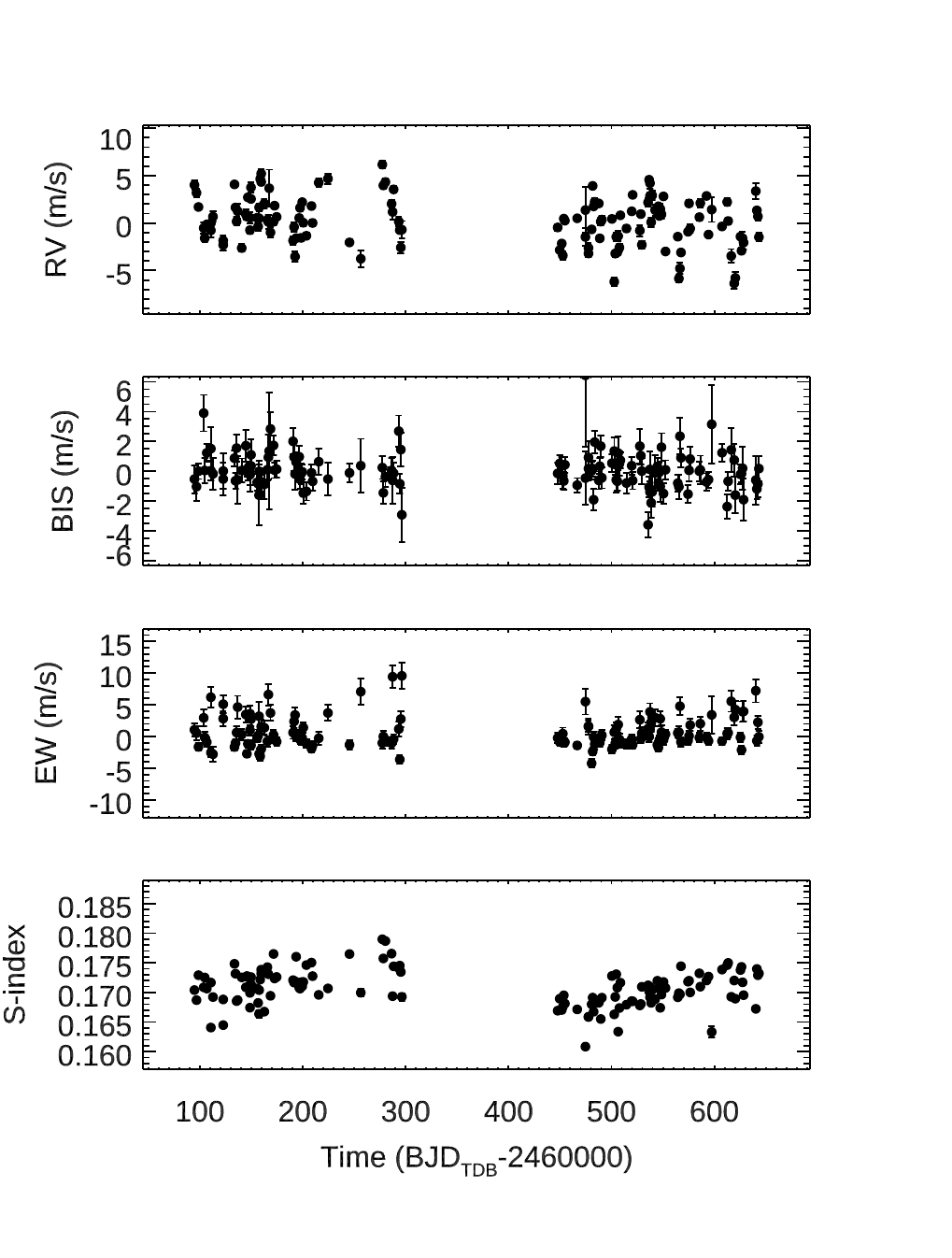}\par
\vspace{-0.5cm}
\caption{Radial velocity (RV) and activity indicators, namely the CCF bisector span (BIS) and equivalent width (EW), and the CaII H\&K $S$-index, as measured from the HARPS-N spectra. The equivalent width is given by the product of the full width of half maximum and the contrast of the CCF.}
\label{fig:RVs_ActInd}
\end{multicols}
\end{figure*}

\begin{table*}
\vspace{2.0 cm}
\centering
\caption{HARPS-N measurements of radial velocity and activity indicators.}
\tiny
\begin{tabular}{c c c c c c c c c c c c c} 
\hline
Time  & RV & $\sigma_{\rm RV}$ & FWHM & $\sigma_{\rm FWHM}$ & BIS & $\sigma_{\rm BIS}$ & C & $\sigma_{\rm C}$ & $S_{\rm MW}$ & $\sigma_{\rm S_{\rm MW}}$ & $\lrhk$ & $\sigma_{\lrhk}$ \\
$[\rm BJD_{TDB}]$ & [\ms] & [\ms] & [\ms] & [\ms] & [\ms] & [\ms] & [\%] & [\%] & & & dex & dex \\
\hline
2460073.656473 &	-49319.612 & 1.457 & 6310.45 &	2.91 &	-67.12	& 2.91 & 63.641 & 0.029 & 0.152 & 0.005 & -5.261 & 0.032 \\
2460073.660514 &	-49312.354 & 1.814 & 6303.35 & 3.63 & -70.60 & 3.63 &	63.684 & 0.037 & 0.143 & 0.007 & -5.329 & 0.053 \\
... & ... & ... & ... & ... & ... & ... & ... & ... & ... & ... & ... & ...\\
\end{tabular}
\tablefoot{From left to right the columns report the epoch of the observation, the radial velocity (RV) and its uncertainty, the activity indicators full width at half maximum (FWHM), bisector span (BIS), and contrast (C) of the cross-correlation function, and the CaII H\&K Mount Wilson $S$-index ($S_{\rm MW}$) and $\lrhk$ indicators, along with their uncertainties. Data are available at both the CDS and DACE databases. A portion is shown here for guidance regarding its form and content. }
\label{tab:data}
\end{table*}
\FloatBarrier

\clearpage

\onecolumn

\section{Data analysis and results}

\subsection{Generalized Lomb Scargle periodograms of the HARPS-N radial velocities}

\begin{figure*}[ht!]
\centering
\vspace{-1cm}
\includegraphics[width=16.5 cm]{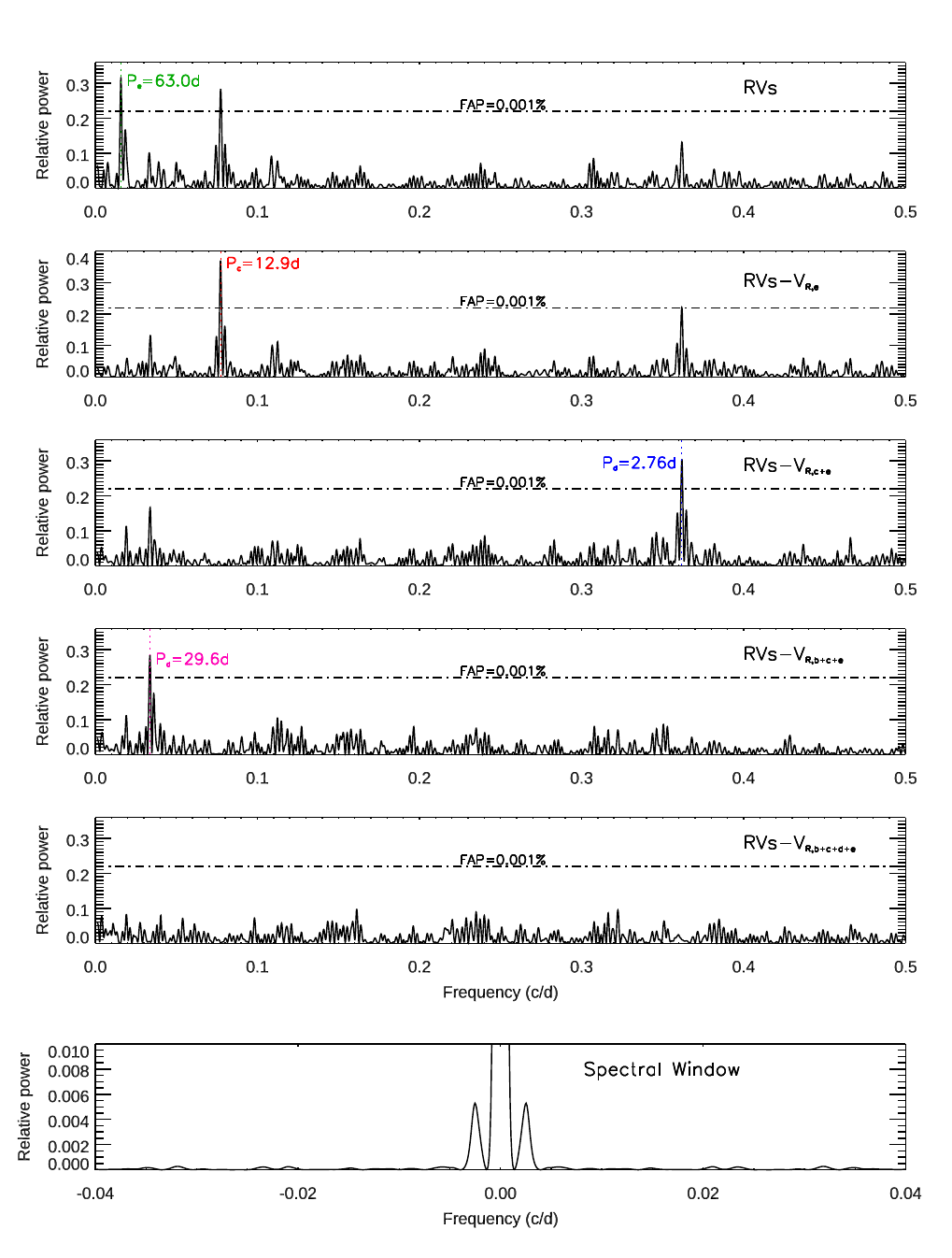}
\caption{Generalized Lomb-Scargle periodograms of the 141 HARPS-N radial velocities and residuals as a function of frequency in cycles/day (or day$^{-1}$). From top to bottom are displayed the periodograms of radial velocities; residuals after subtracting the signal of TOI-5789\,e; residuals after subtracting the signals of both TOI-5789\,c and e; residuals after subtracting the signals of TOI-5789\,b, c, and e; residuals after subtracting the signals of TOI-5789\,b, c, d, and e; the spectral window showing the most prominent lobes of the yearly alias. The dotted vertical lines indicate the orbital periods of planets e (green), c (red), b (blue), and d (pink). The horizontal dash-dotted lines show the false alarm probability level of $10^{-5}$.}
\label{fig:GLS_periodograms}
\end{figure*}

\twocolumn


\begin{figure}[ht!]
    \centering
\vspace{-0.5 cm}
\includegraphics[width=0.525\textwidth]{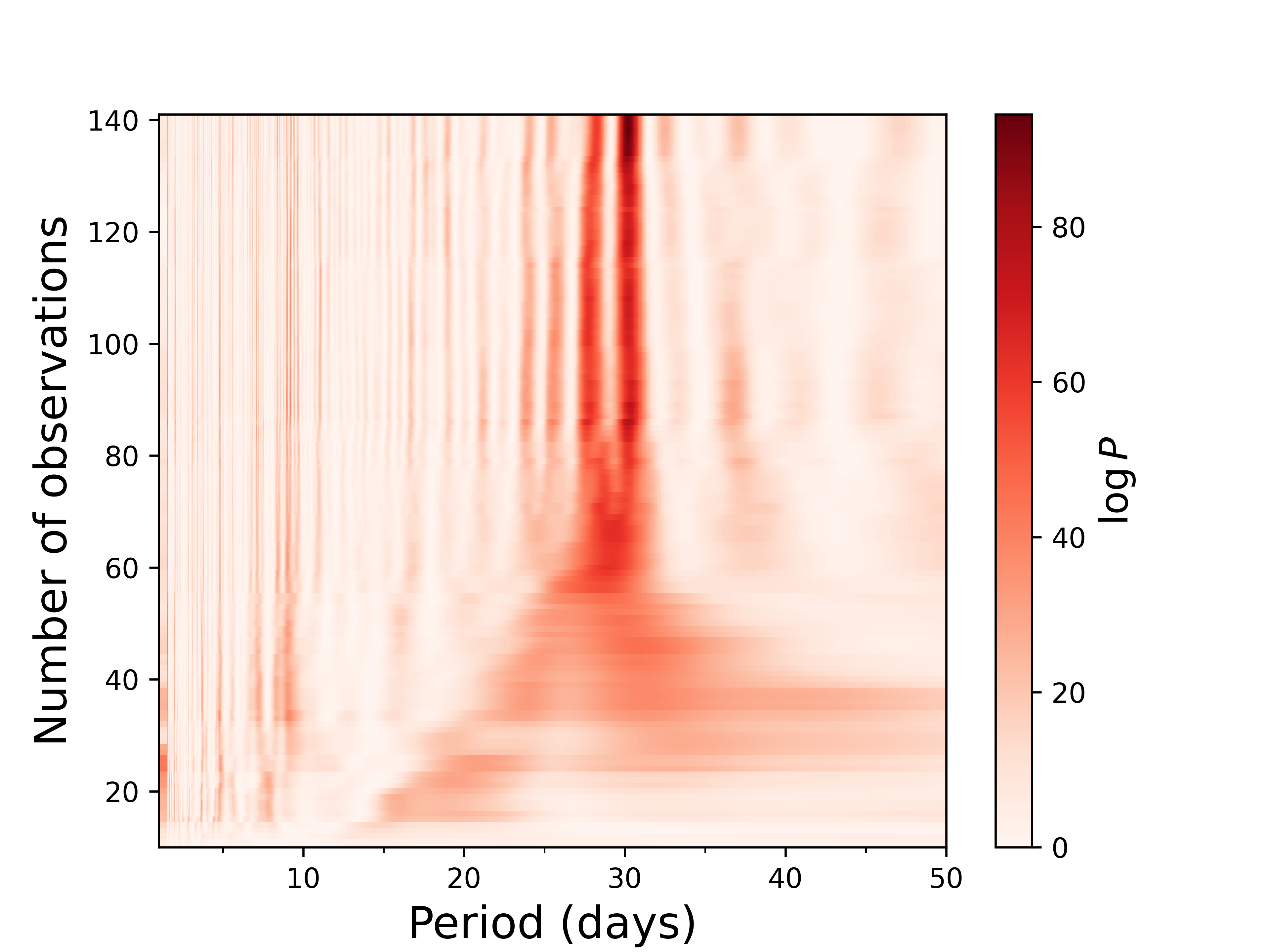}
    \caption{Stacked Bayesian Generalized Lomb Scargle periodogram of the HARPS-N radial-velocity residuals after subtracting the signals of planets b, c, and e through a recursive pre-whitening (see Fig.~\ref{fig:GLS_periodograms}).}
    \label{fig:sbgls}
\end{figure}

\subsection{Apodized sine periodograms}
\label{app:asp}

 To compute the apodized sine periodogram (ASP), 
 we considered a grid of timescales $\tau = 10\cdot T_{\rm obs}, T_{\rm obs}/3, T_{\rm obs}/9$ and $T_{\rm obs}/27$, where $T_{\rm obs}$ is the total observation timespan of the data. We compared the $\chi^2$ of a linear base model $H$, and the $\chi^2$ of a model $K(\omega, t_0,\tau)$ 
defined as the linear model of $H$ plus an apodized sinusoid $e^{-\frac{(t-t_0)^2}{2\tau^2}}  (A \cos \omega t + B \sin \omega t )$.
The ASP is defined as the $\chi^2$ difference between the two models
\begin{align}
	z(\omega, t_0,\tau) &= \chi^2_H - \chi^2_{K(\omega, t_0,\tau)}.\label{eq:z}
\end{align}

\noindent
At each frequency, we fit a linear model containing an offset, the FWHM, bisector span, and $S$-index. 
We considered the HARPS-N data only, and assumed an uncorrelated noise model, with a jitter of 1 m/s. 
 
\begin{figure*}
	\noindent
	\centering
	\hspace{-1,5cm}
	\begin{tikzpicture}		
\path (0,0) node[above right]{\includegraphics[width=0.8\linewidth]{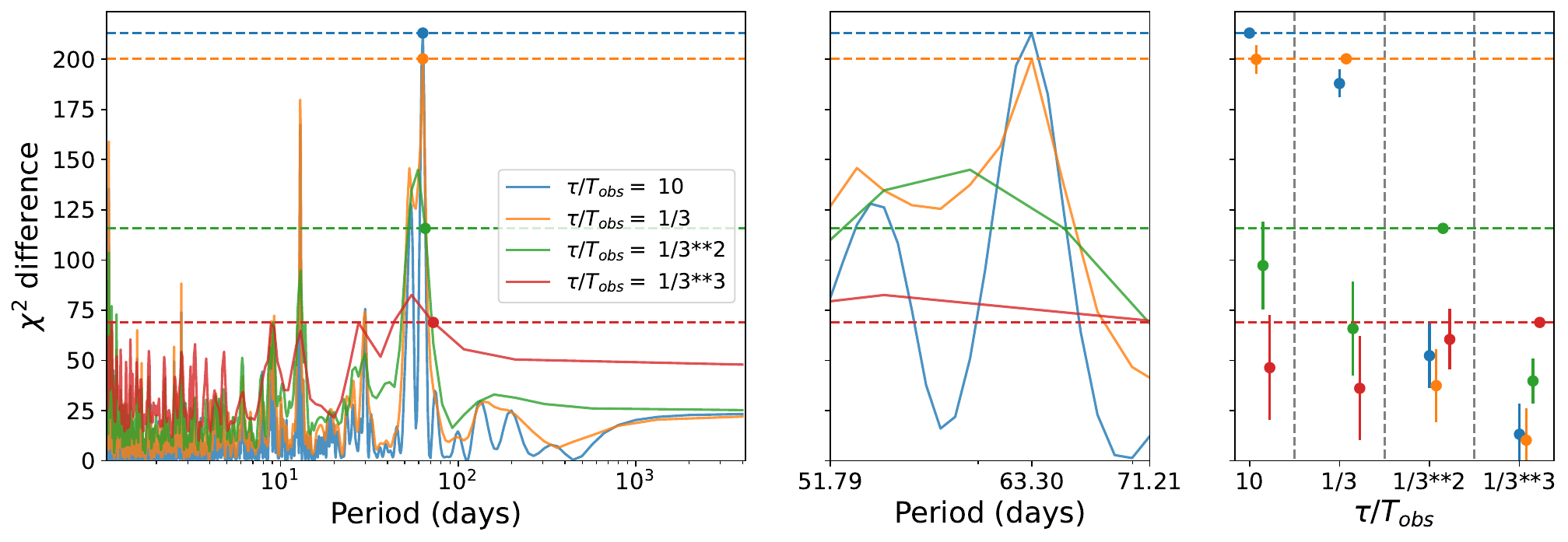}};
	\path (1.15,4.5) node[above right]{\large(a)};
	\begin{scope}[yshift=-5.5cm]
\path (0,0) node[above right]{\includegraphics[width=0.8\linewidth]{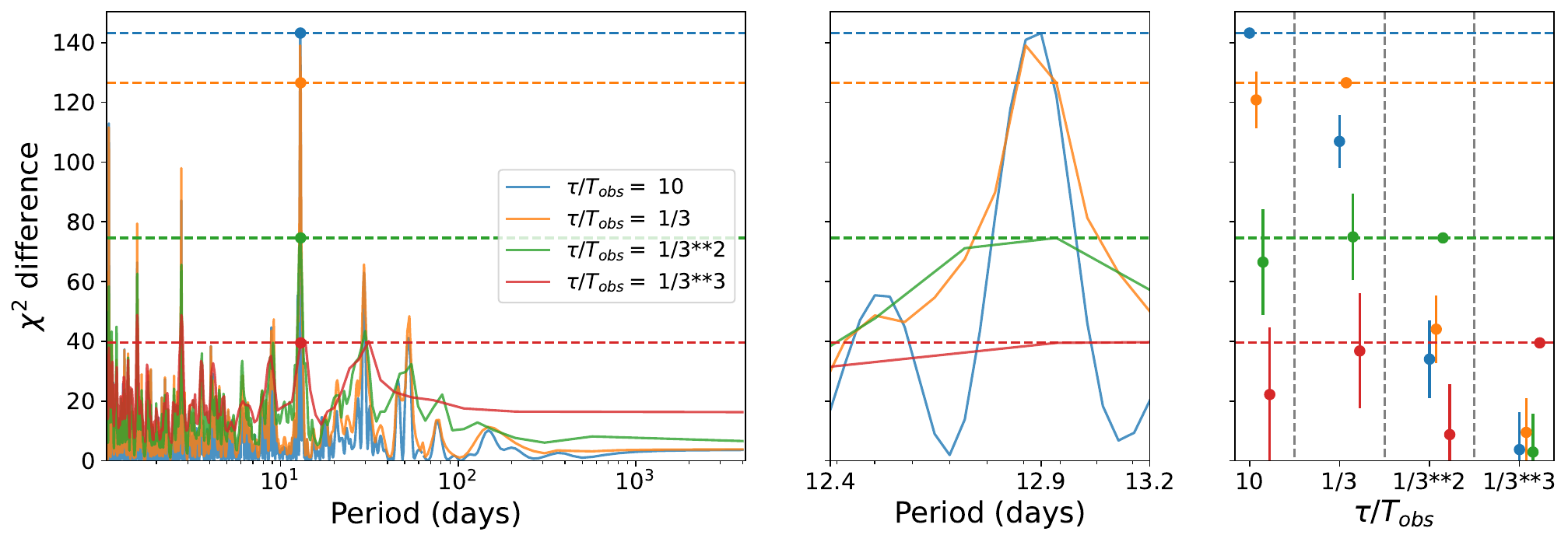}};
	\path (1.15,4.5) node[above right]{\large(b)};
	\end{scope}
	\begin{scope}[yshift=-11cm]
\path (0,0) node[above right]{\includegraphics[width=0.8\linewidth]{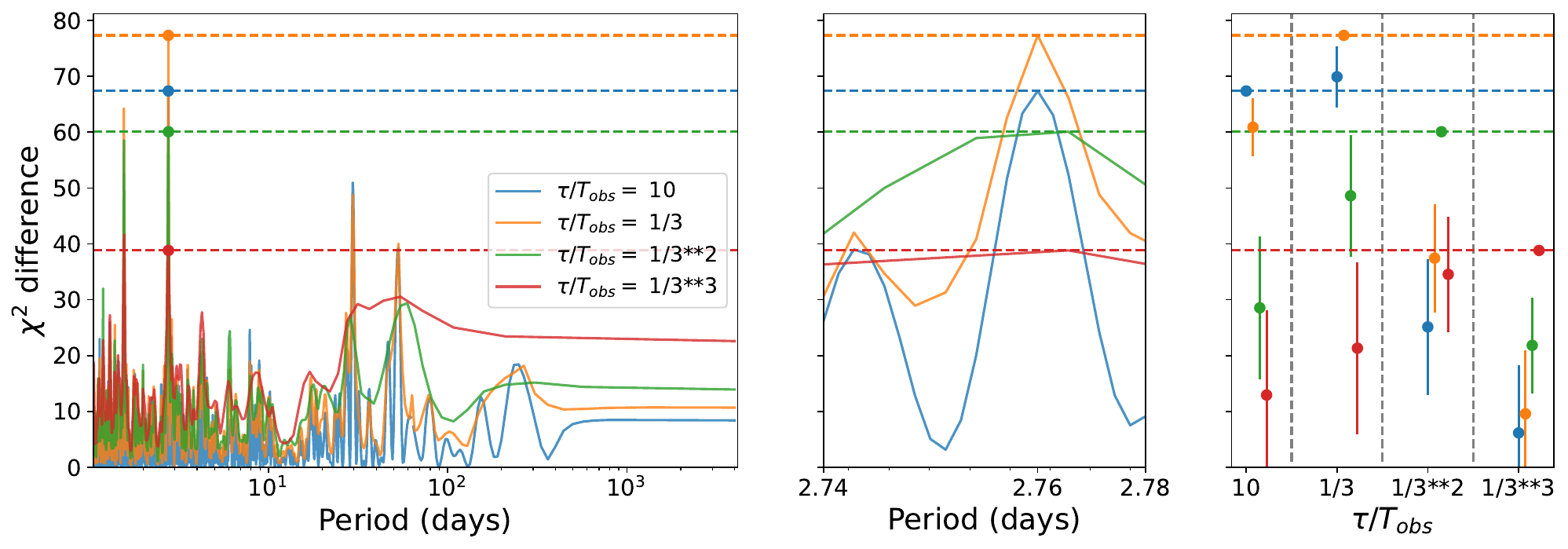}};
	\path (1.15,4.5) node[above right]{\large(c)};
	\end{scope}
	\begin{scope}[yshift=-16.5cm]
\path (0,0) node[above right]{\includegraphics[width=0.8\linewidth]{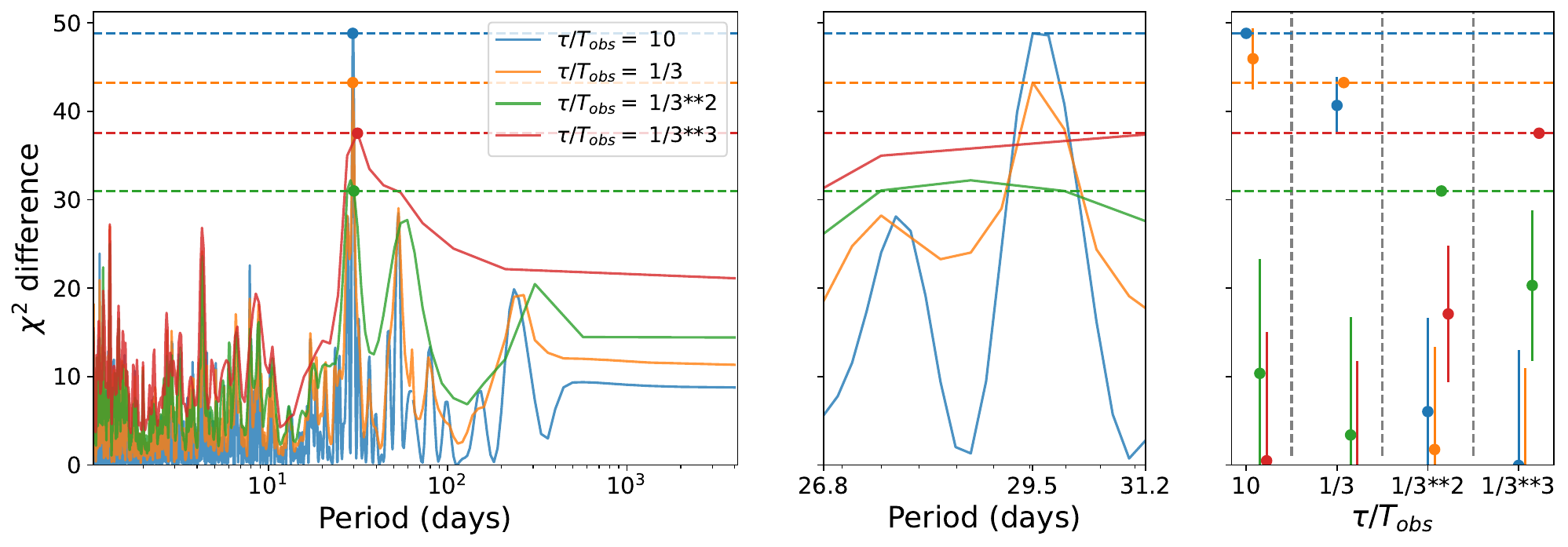}};
	\path (1.15,4.5) node[above right]{\large(d)};
	\end{scope}
	\end{tikzpicture}
	\caption{Four first iterations of the apodized sine periodogram (ASP) method. Models correspond to the maximum of the ASPs. 
	The left panels display the periodograms, the middle panels a zoom in on the highest peak, 
 	and the right hand panel represents a statistical significance test (see text for details).}
	\label{fig:ASP}
\end{figure*}
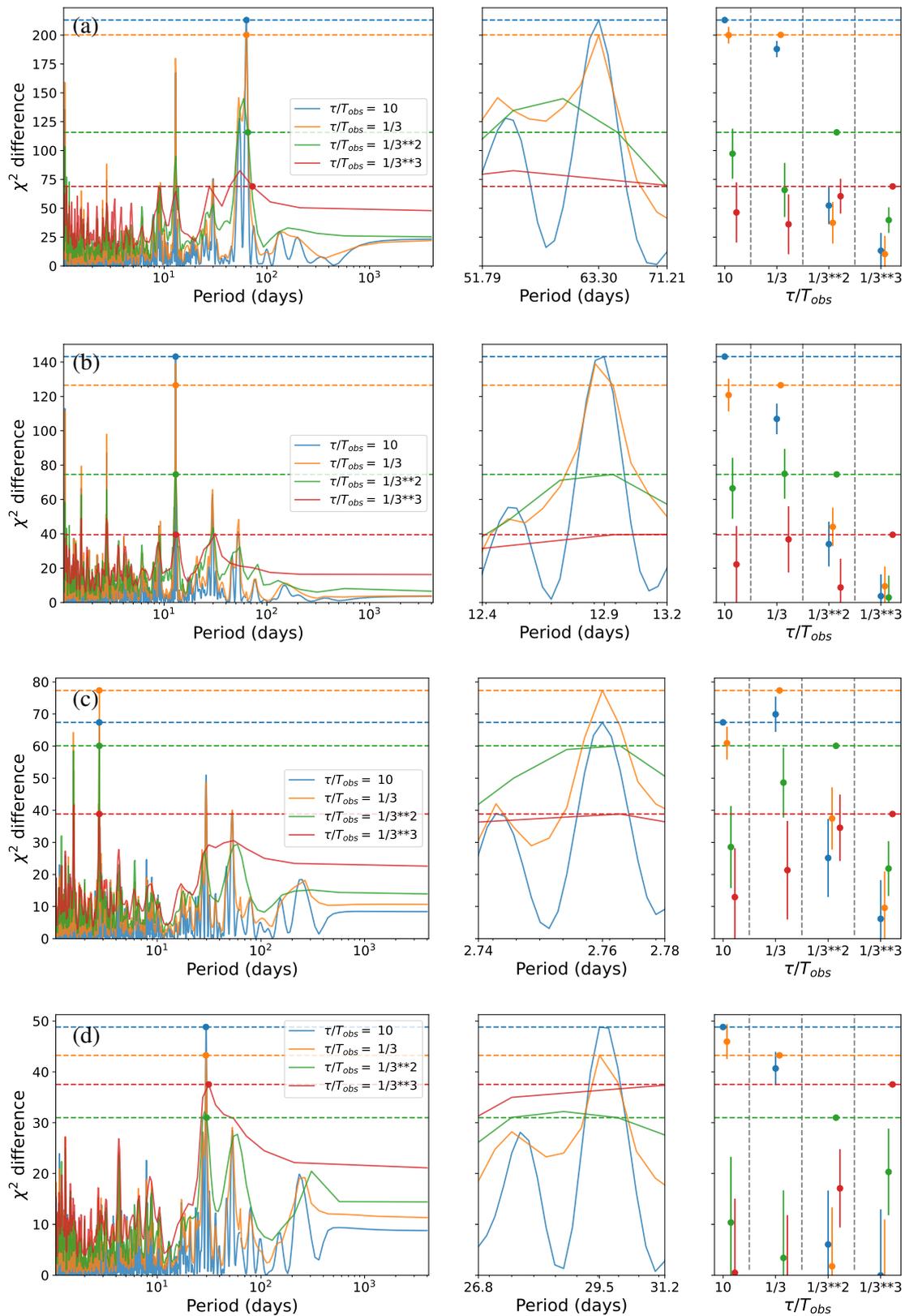

In Fig. \ref{fig:ASP}.a, left panel, we represent $z(\omega, t_0,\tau)$  as defined in Eq.~\eqref{eq:z}, maximized over $t_0$ for the values of $\tau$ in the grid. 
 In other words, we always selected the center of the time window which best fits the data. We searched iteratively for signals. Once the maximum peak of the ASP was found, 
 we then included the corresponding model in the base linear model $H$, and repeated the process. 
 The four first iterations are shown in Fig.~\ref{fig:ASP}, a, b, c, and d. 
 The left panels display the periodograms, the middle panels a zoom in on the highest peak, 
 and the right hand panel represents a statistical significance test, as explained below. 
 
In Fig.~\ref{fig:ASP}.a, the dominating peak has a period of 63.3 d. 
We want to test whether the signal is statistically compatible with a constant one (i.e., with $\tau = 10\cdot T_{\rm obs}$). 
Denoting by $t_{(\tau,\omega)}$ the value of $t_0$ maximizing the value of the periodogram~\eqref{eq:z} for a given $\omega$ and $\tau$,  
we compute the distribution of 
\begin{align}
   D_z = z(\omega, t_{(\tau,\omega)}, \tau) - z(\omega, t_{(\tau^\prime,\omega)}, \tau^\prime) 
   \label{eq:dz}
\end{align}
 with the hypothesis that the model $K(\omega, t_{(\tau,\omega)}, \tau, A^\star, B^\star )$ is correct, 
 where the fit cosine and sine amplitudes $A^\star, B^\star$ are obtained by fitting model $K$ to the data. $D_z$ can easily be expressed as a generalized $\chi^2$ distribution, 
 with mean and variance given by an analytical expression \citep{hara2022b}. 

 In Fig.~\ref{fig:ASP}.a, c and d we can see that for all three signals at 63, $12.9$, and $30$\,d, the preferred timescale is the longest one. As for the $2.76$ signal,  it is still compatible with a purely periodic one. 
 As a remark, if we use a correlated noise model instead of simply an additional jitter, the preferred timescale for the 30\,d signal becomes shorter. Nonetheless, we do not consider this a sufficient reason to put in serious doubt the planetary origin of the signal, given that significant correlated noise due to stellar activity is not expected for such an old and inactive star.

\clearpage

\onecolumn

\subsection{Bayesian priors}

\begin{table}[ht!]
\vspace{1.0 cm}
\centering 
\caption{Priors imposed in the DE-MCMC radial-velocity and/or transit+radial-velocity combined analyses.} 
\vspace{1.0 cm}
\begin{tabular}{l c | l c}
\hline
\hline
\multicolumn{4}{c}{\textbf{\large{Priors on radial-velocity parameters}}} \\ 
\hline
\hline
\multicolumn{2}{c |}{\textbf{\large{3-planet+GP model}}} & \multicolumn{2}{c}{\textbf{\large{4-planet model}}} \\ 
\hline
$\gamma$ [\ms] & $U]-\infty, +\infty[$ & $\gamma$ [\ms] & $U]-\infty, +\infty[$ \\
$\sigma_{\rm RV,jit}$ [\ms] & $U[0, +\infty[$ & $\sigma_{\rm RV,jit}$ [\ms] & $U[0, +\infty[$ \\
$h$ [\ms] & $U[0, +\infty[$ & & \\
$\lambda_1$ [days] & $U[0, 1000]$  & & \\
$\lambda_2$ & $U[0.1, 10]$  & & \\
$P_{\rm rot}$ [days] & $U[27, 33]$  & & \\
\hline
\multicolumn{2}{c |}{TOI-5789\,b} & \multicolumn{2}{c}{TOI-5789\,b}  \\
\hline
$T_{\rm c} \rm [BJD_{TDB}-2450000]$ & $\mathcal{U}[10357.1, 10359.1]$ & $T_{\rm c} \rm [BJD_{TDB}-2450000]$ & $\mathcal{U}[10357.1, 10359.1]$  \\
$P$ [days] & $\mathcal{U}[2.74, 2.78]$ & $P$ [days] & $\mathcal{U}[2.74, 2.78]$ \\
$e$ & $\rm 0~(fixed)$ & $e$ & $\rm 0~(fixed)$ \\
$K$ [\ms] & $U[0, +\infty[$ & $K$ [\ms] & $U[0, +\infty[$ \\
\hline
\multicolumn{2}{c |}{TOI-5789\,c} & \multicolumn{2}{c}{TOI-5789\,c} \\
\hline
$T_{\rm c} \rm [BJD_{TDB}-2450000]$ & $\mathcal{N}(10151.15868, 4.9\mbox{\sc{e}-04})$ & $T_{\rm c} \rm [BJD_{TDB}-2450000]$ & $\mathcal{N}(10151.15868, 4.9\mbox{\sc{e}-04})$ \\
$P$ [days] & $\mathcal{N}(12.927748, 1.6\mbox{\sc{e}-05})$ & $P$ [days] & $\mathcal{N}(12.927748, 1.6\mbox{\sc{e}-05})$ \\
$e$ & $\mathcal{N}(0.0, 0.098)~\rm AND ~ 0 \leq e < 1$ & $e$ & $\mathcal{N}(0.0, 0.098)~\rm AND ~ 0 \leq e < 1$ \\
$K$ [\ms] & $U[0, +\infty[$ & $K$ [\ms] & $U[0, +\infty[$ \\
\hline
\multicolumn{2}{c |}{ } & \multicolumn{2}{c}{TOI-5789\,d} \\
\hline
 & & $T_{\rm c} \rm [BJD_{TDB}-2450000]$ & $U[10342, 10352]$ \\
 & & $P$ [days] & $U[27, 33]$  \\
 & & $e$  &  $ \mathcal{N}(0.0, 0.098)~\rm AND ~ 0 \leq e < 1 $ \\
 & & $K$ [\ms] & $U[0, +\infty[$  \\
\hline
\multicolumn{2}{c |}{TOI-5789\,e} & \multicolumn{2}{c }{TOI-5789\,e}\\
\hline
$T_{\rm c} \rm [BJD_{TDB}-2450000]$ & $U[10350, 10370]$ & $T_{\rm c} \rm [BJD_{TDB}-2450000]$ & $U[10350, 10370]$ \\
$P$ [days] & $U[60, 66]$ &  $P$ [days] & $U[60, 66]$ \\
$e$  &  $ \mathcal{N}(0.0, 0.098)~\rm AND ~ 0 \leq e < 1 $ & $e$  &  $ \mathcal{N}(0.0, 0.098)~\rm AND ~ 0 \leq e < 1 $ \\
$K$ [\ms] & $U[0, +\infty[$ & $K$ [\ms] & $U[0, +\infty[$ \\
\hline
\hline
\end{tabular}
\begin{tabular}{l l c r}
\multicolumn{4}{c}{\textbf{\large{Priors on TOI-5789\,c transit parameters}}} \\ 
\hline
\hline
& $\sigma_{\rm phot, s54}$ [rel. flux] & $U[0, +\infty[$ & \\
& $\sigma_{\rm phot, s82}$ [rel. flux] & $U[0, +\infty[$ & \\
& $\rho_\star$ [g\,cm$^{-3}$] & $\mathcal{N}(2.00,0.17)$ & \\
& $T_{\rm c} \rm [BJD_{TDB}-2450000]$ & $\mathcal{U}]-\infty, +\infty[$ & \\
& $P$ [days] & $\mathcal{U}[0, +\infty[$ \\
& $R_{\rm p}/R_{*}$ & $U[0, +\infty[$ & \\ 
& $T_{14}$ [d] & $U[0, +\infty[$ & \\
& $i$ [deg] & $U[0, 90]$ & \\
& $q_1$  & $U[0, 1]$ & \\
& $q_2$  & $U[0, 1]$ & \\
\hline
\hline
\label{tab:priors_parameters}
\end{tabular}
\end{table}
\tablefoot{Same parameters as in Table~\ref{tab:planet_parameters}. We note that $\mathcal{N}$ and $U$ stand for normal (Gaussian) and uniform priors, respectively.}



\onecolumn

\noindent \begin{minipage}{\textwidth}
\section{Composition of TOI-5789\,c}
\begin{minipage}[t][6cm][t]{0.49\textwidth}
\subsection{Priors and posteriors}
\begin{table}[H]
\centering
\caption{Prior parameter distribution for the calculation of the internal compositions of TOI-5789\,c, where $M_\mathrm{atm}$ is the atmosphere mass, $Z_\mathrm{env}$ the envelope water mass fraction, M$_\mathrm{core+mantle}$ the mass of the deep interior including core and mantle, for which we assume an Earth-like core-to-mantle ratio of 0.325:0.675. The prior on $M_\mathrm{atm}$ is equivalent to mass fractions between 0.5 - 15\%. }\label{tab:comppriors}
\renewcommand{\arraystretch}{1.2}
\tiny
\begin{tabular}{lcc}
    \hline\hline
     Parameter & Prior range & Distribution \\
    \hline \\[-6pt]%
    $M_\mathrm{atm}/M_{\oplus}$ & $0.025-0.75$ & log-uniform \\
    $M_\mathrm{core+mantle}$ & $\mathcal{N}(M_{\rm p}, \sigma^2_{M_{\rm p}})$  & gaussian  \\
    $Z_\mathrm{env}$ (envelope water mass fraction) & $0.02-1.0$ & uniform\\
    \bottomrule
\end{tabular}
\end{table}
\end{minipage}
\hspace{0.02\textwidth}
\begin{minipage}[t][6cm][t]{0.49\textwidth}
\subsection{Accuracy of the surrogate model}
For this inference, the surrogate model provides high quality fits with R-squared values (coefficient of determination) of 0.9999 and 0.9999 for the planetary mass and radius, respectively. Also, the surrogate root mean square errors are well below observational uncertainties with 0.0004 and 0.0028 for planetary mass and radius, respectively. Those errors of the model uncertainty are accounted for in the likelihood function.
\end{minipage}
\end{minipage}

\begin{figure*}[h!]
\vspace{1cm}
    \centering
\includegraphics[width=0.75\linewidth]{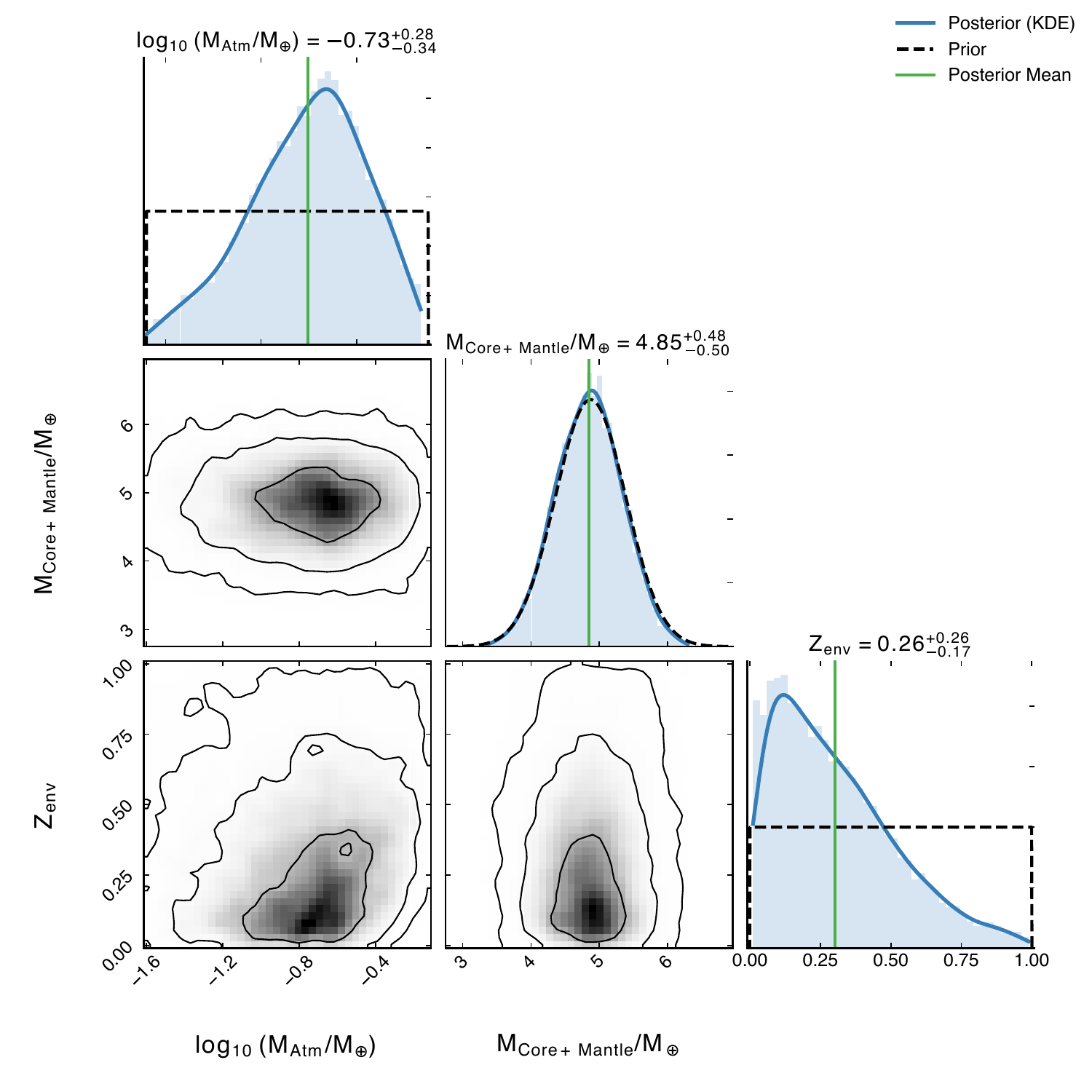}
    \caption{1D and 2D marginalized posterior distributions for the internal composition analyses of TOI-5789\,c.}
    \label{fig:composition}
\end{figure*}
\FloatBarrier

\clearpage
\newpage

\onecolumn

\section{Prospects for atmospheric characterization ot TOI-5789\,c}

\begin{figure*}[ht!]
\vspace{2cm}
\centering
\includegraphics[width=0.875\textwidth]{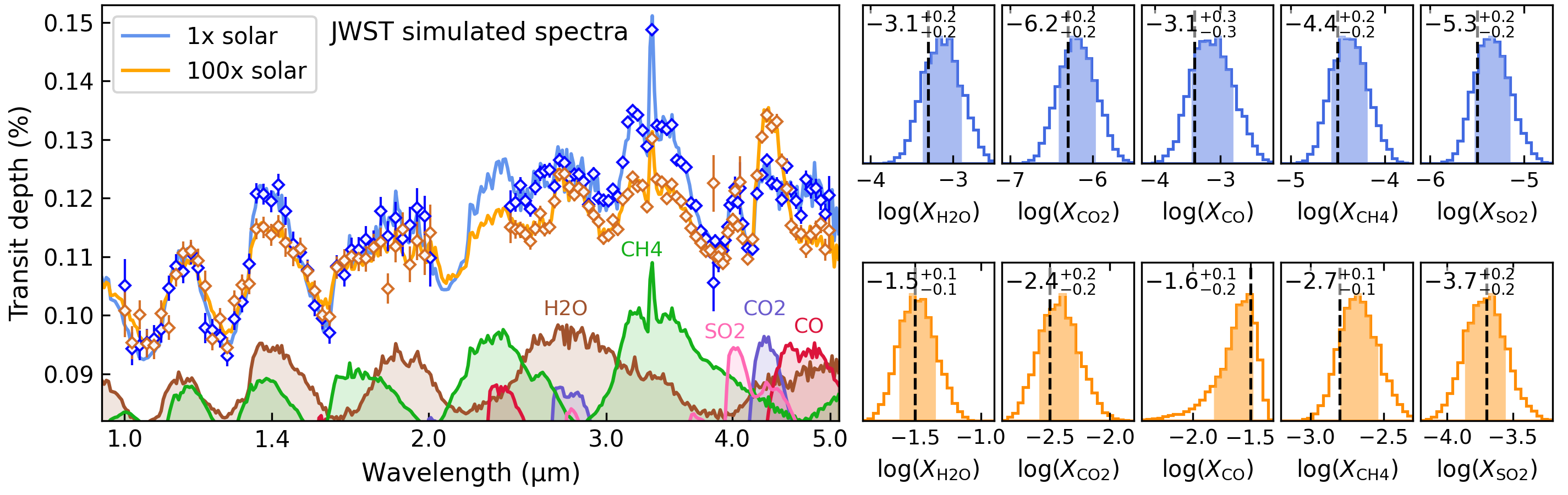}
\caption{Left: Simulated JWST transmission spectra of TOI-5789\,c (top panels) for atmosphere scenarios at $1\times$ (blue) and $100\times$ (orange) solar metallicities. The diamond markers with error bars denote the simulated NIRCam observations, which probe the 1.0--5.0 {\micron} range (F150W2, F322W2, and F444W). The lower part of the panel shows the scaled contributions of different molecules. Right: Retrieved posterior distributions for the VMRs of the molecules expected to shape the transmission spectrum (same color coding as before). The shaded areas denote the 1$\sigma$ uncertainty. The values report the median and 1$\sigma$ span of the posteriors. The black dashed lines denote the VMRs of the underlying true models.}
\label{fig:atmospheric_retrievals}
\end{figure*}

\begin{figure*}[ht!]
\centering
\vspace{2cm}
\includegraphics[width=0.875\textwidth]{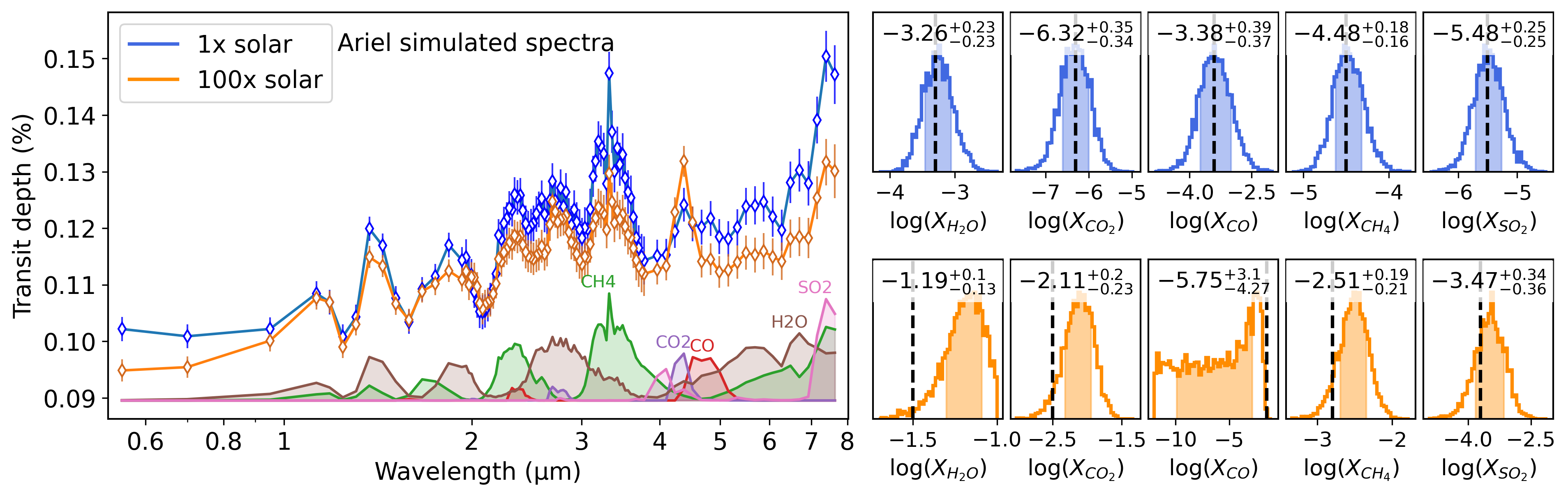}
\caption{Left: Simulated spectra of TOI-5789 at Ariel’s Tier-3 resolution, assuming 1× (blue) and 100× (orange) solar metallicities. The lower part of the panel shows the scaled contributions of different molecules. Right: Retrieved chemical abundances. The dashed vertical lines indicate the true values for each species, while the colored regions of the histograms represent the 15.86\%-84.14\% quantiles of the posterior distributions. }
\label{fig:ariel_atmospheric_retrievals}
\end{figure*}

\end{document}